\documentclass[11pt,a4paper]{article}
\usepackage{t1enc}
\usepackage[latin1]{inputenc}
\usepackage[english]{babel}
\usepackage{harvard}
\usepackage{graphicx}

\newcommand{\R}{\mbox{\rm I\hspace{-0.33ex}R}}

\newcommand{\beq}{\begin{equation}}
\newcommand{\eeq}{\end{equation}}
\newcommand{\beqarr}{\begin{eqnarray}}
\newcommand{\eeqarr}{\end{eqnarray}}
\newcommand{\beqa}{\begin{eqnarray*}}
\newcommand{\eeqa}{\end{eqnarray*}}
\newtheorem{theorem}{Theorem}
\newtheorem{theorem(Weyl)}{Theorem (Weyl)}
\newtheorem{lemma(Weyl)}{Lemma (Weyl)}
\newtheorem{proposition}{Proposition}
\newtheorem{lemma}{Lemma}

\newtheorem{remark}{Remark}
\newtheorem{definition}{Definition}

\newtheorem{principle}{Principle}

\newtheorem{terminology}{Terminology}
\newtheorem{example}{Example}

\begin{document}
\begin{center} 
\thispagestyle{empty}
\subsection*{An extended frame for cosmology  by integrable\\ Weyl geometry }
{\em Erhard Scholz,\footnote{scholz@math.uni-wuppertal.de}\\
Wuppertal University\\
Department C, 
 Div. of Mathematics\\ 
 42097 Wuppertal, Germany }   \\[2.5pt]
\end{center}

\begin{abstract}
A conservative extension of general relativity by integrable Weyl geometry is formulated, and a new class of cosmological models ({\em Weyl universes}) is introduced and studied. A short discussion of how these new models behave in the light of observational evidence follows. It turns out that the subclass of 
Weyl universes  with positive ex-ante curvature of spatial sections ({\em Einstein-Weyl universes}) forms an empirically contentful conceptual frame for cosmology. Recent high precision data   (supernovae Ia, quasar frequencies,  microwave background) are well represented in this type of Weyl geometry and  several of the anomalies of the standard approach are avoided or   become meaningless.
\end{abstract}

\subsubsection*{Introduction}
Present standard cosmology relies on  the physical  geometry of 
Friedman-Lema\^itre models and draws upon the hypothesis of  epanding spatial sections in relativistic space-time. 
For several decades, this approach was a satisfying conceptual frame for astrophysical and astronomical 
research. Even though many   continue to consider it as a reliable guide
 for the evaluation of empirical observations, there are strong indications of growing anomalies
for this interpretational scheme, most prominent among them the appearance of  
``non-baryonic dark matter'', 
the surprising ``time-dependence of vacuum (dark) energy'') and the ``scale problem of vacuum energy''.
Other problems, like the  isotropy/anisotropy riddle of the cosmic microwave background, are 
covered  up by surprising towers of ad-hoc constructions or are neglected, like the negative results on a systematic correlation between redshift and metallicity of galaxies and quasars. It is not the  goal of this paper to discuss 
these questions in detail.  We rather want to show that already a small and innocent looking modification in the conceptual framework of 
 Robertson-Walker manifolds which underly the  geometry of the Friedman-Lema\^itre 
models (F-L), allows to conceive of  new physical geometries which are consistent with general relativity (GRT).\footnote{For a general survey of non-Riemannian geometries as possible alternatives
 for cosmological geometry see \cite{Puetz:Survey}.}
 They can be well adapted to observational data  and give an account  of several 
of the crucial tests for cosmological 
models,  at least as convincingly as in the F-L approach (redshift-distance relationship, SN$_{Ia}$ magnitudes).
In some respects, they are  even  more convincing (redshift dependence of quasar frequencies and metallicity), in others new possibilities are open for research (microwave background). Above all, no dynamical ``dark energy'' anomaly appears in these models. The right hand side of the Einstein equation looks much more realistic than in the expanding world approach, although also here questions with respect to the  state equation of the cosmic medium remain.

Since the 1920s, when cosmological redshift started to be observed and became the first empirical 
clue to the construction of cosmological  geometries backed on general relativity, some of the leading protagonists, including
E. Hubble himself, could never  be convinced that the space-kinematical interpretation of  cosmological redshift was more than a 
provisonal move to characterize    the 
energy reduction  of photons  transmitted over long cosmic distances, while in the future a ``more physical explanation'' (H. Weyl in 1930) might be found. Different proposals for  ``tired
 light'' approaches, as   they have  been called by their critics at first,  have been tried, among them the ideas  of  a ``gravitational drag'' of photons (F. Zwicky, MacMillan e.a.) or of a quantum mechanical  photon-photon interaction  (E. Finlay-Freundlich, M. Born, J.-P. Vigier e.a.).\footnote{For Finlay-Freundlich see \cite[634ff.]{Hentschel:Rotverschiebung} and \cite[187ff.]{Hentschel:Freundlich} or  the less critical report (including remarks on Born) \cite{Assis:Redshift}.}
 Although
these investigations  have not  acquired  a physically reliable, in some sense  ``verified'' status, they contributed to a first exploration of possible physical causes of  such a hypothetical mechanism.  
 With the presently maturizing crisis in standard cosmology, it seems more than justified to understand better, how such hypothetical explanations can be conceptually integrated  into general relativistic cosmology  and to explore their  mathematical relationship with the  standard approach. The result of such investigations may  contribute to a change of the symbolic  ``a-priori'' (in a technical sense) of the evaluation of empirical data in observational cosmology. If not, they should at least add to our  understanding of the  expanding space cosmologies from ``outside'', i.e., without  assuming an expansion as an ex-ante property of the models.

In this article, we analyze cosmological redshift in Robertson-Walker models in terms of 
(scale-integrable) Weyl geometry. This allows to refrain from any ex-ante decision as to  the physical explanation of cosmological redshift,  either  as space-kinematical (``Doppler'') or as tired light. It even allows us, to a certain degree, to  translate mathematically between the two. We will see that there are not only those Robertson-Walker models which have a  convincing  physical interpretation   under the assumption of a ``real'' space-expansion (the classical 
Friedman-Lema\^itre models), but also others which allow a convincing physical  interpretation  under the assumption of a tired light hypothesis. There is a particularly simple class of physically meaningful   Robertson-Walker models of the new type, which will be called {\em Weyl universes}, because they rely  esssentially on simple aspects of Weyl geometry.\footnote{They were {\em not} already introduced by H. Weyl, although he could easily have done so, had he broken with the dogma of the electromagnetic potential of his gauge geometry early enough.}
  It is  possible   to interpret the data of observational cosmology in both model classes and to see in which approach  physico-geometrical reality appears  more convincing. That cannot be done  here in detail, although some indications on first results of empirical evaluations (``tests'' of the models) will be given at the end  of this paper.

\subsubsection*{1. Weyl geometry }

H. Weyl's  generalization of Riemannian and semi-Riemannian geometry, first presented in \cite{Weyl:InfGeo},  is still being discussed in the actual literature, although the original intentions of its inventor for its physical application have not been realized. For  
mathematicians, Weyl geometry became an interesting structure after fibre bundles had been invented and Weyl's  idea could be rephrased in this language \cite{Folland:WeylMfs,Varadarajan:Connections}. In the physical literature it played an important part  in the first wave of unified field theories and  seemed to  loose its importance with it.\footnote{See  \cite{Goenner:UFT}.}
But it contained much broader  possibilities  than its direct semantical link to Weyl's first (metrical) gauge theory of electromagnetism  made visible.   In his  ``mathematical analysis of the space problem'' during the years 1921--23,   Weyl gave strong arguments  in this direction.\footnote{See \cite{Scholz:TelAviv}.}
The
conceptual potential   was  taken up again, in a different sense, by J.  Ehlers, F. Pirani and A. Schild in  their constructive axiomatic approach to  general relativity 
\cite{Ehlers/Pirani/Schild}. Thus the question arose, whether the additional structural element of Weyl geometry, the length connection $\varphi$,  could play the role of some, perhaps still unknown, field. A negative answer  was given by   J. Audretsch,  F. G\"ahler and N. Straumann, who  showed that a Weyl geometry can only  be considered as compatible with relativistic quantum mechanics, if the length connection is integrable \cite{Audretsch/Gaehler/Straumann}. Such a restriction had already been proposed by P.A.M. Dirac in his \cite{Dirac:1973}.  Field theoretically,  the extension of GRT by Weyl geometry   appeared thus to be condemned to triviality.

This is not the end of the story, however; see, e.g., \cite{Tiwari:Weyl_geometry,Tiwari:Einstein_Weyl}. 
Here we reconsider Weyl geometry from a  genuinely geometrical perspective  which is,
 in  its metrical aspects, very close to 
Weyl's original view. In this section   the aspects of Weyl geometry which are most important for us will be  resumed.\footnote{Although there is considerable overlap with section 3 of \cite{Dirac:1973}, readers acquainted with Dirac's formalism should be aware that Dirac switched sign in the transformation of the length connection, in our notation $\tilde{\varphi} = \varphi + d \ln \lambda$ rather than  $\tilde{\varphi} = \varphi - d \ln \lambda$ \cite[equs. (1.2), (1.3)]{Dirac:1973}]. This change destroys the behaviour as a proper gauge transformation in the sense of differential geometry (footnote \ref{fn gauge transformation}). In particular, Weyl's gauge invariant comparison of length ratios (our lemma 1) gets lost. \label{fnDirac1} }
 In the next one we see  how it can be 
used for a weak, i.e. conservative, extension  of general relativity, even if the restriction of Audretsch, G\"ahler and Straumann is taken into account. In the third and the following sections,   a method is developed which allows  to use integrable Weyl geometry for a generalized view on Robertson-Walker manifolds and for cosmological model building.

Weyl proposed to found the metrical concepts of differential geometry  on a ``purely infinitesimal'' point of view more strictly than in Riemannian geometry. In particular, he was not satisfied with the possibility,   inbuilt  into classical differential geometry, to  compare   lengths of vectors at different points of a manifold  directly. He  considered it as more  natural to compare  lengths of vectors directly, only if they are attached to the same point, $\xi, \eta \in T_pM$ for some $p \in M$. That can   easily  be realized  by  a specification of   a conformal 
semi-Riemannian structure on $M$. For the comparison of lengths of vectors at different points $p, q \in M$, $\xi \in T_pM, \, \eta   \in T_qM$, Weyl    introduced   a  {\em length connection}, given by a differential 1-form $\varphi$ on $M$. This form was used to define a kind of  ``length transfer''  by integration along paths, comparable to the parallel transfer of vectors by an affine connection. Of course, other geometrical or physical quantities related to length or time measurement (``gauge quantities'' here introduced  as Weyl fields  in definition  5)  have to be compared in a similar manner. The change of the representative of the conformal metric leads to a change of the differential 1-form of the length connection \cite[122ff.]{Weyl:RZM5}. The result of Weyl's rearrangement of basic concepts of differential geometry can be stated in the following terms:\footnote{The formulations of this passage owes much to proposals by M. Kreck. Compare also \cite{Folland:WeylMfs,Varadarajan:Connections}.}

\begin{definition}
a)  A {\bf  gauge} of a  differentiable manifold $M$  is  given by a pair $(g, \varphi)$ consisting of a semi-Riemannian metric $g$ on $M$ of any signature $(p,q)$  (here usually of Lorentzian signature $ (p,q) = (1,3) )$ and a real- valued differential 1-form $\varphi$ on $M$. The 1-form $\varphi$ will be called the {\bf length connection} (or {\bf scale connection}) of the gauge and $g$ its {\bf Riemannian component}.\\[0.5pt]
b) Two gauges  $(g, \varphi)$ and  $(\tilde{g}, \tilde{\varphi})$ are considered as {\bf gauge equivalent}, iff there is a strictly positive real valued function $ \lambda $ on $M$, such that
\begin{eqnarray} 
 \label{gauge-factor} \tilde{g} &=& \lambda ^2 g , \\ 
 \label{gauge-transformation} \mbox{and}\;\;\;  \tilde{\varphi} &=& \varphi -  d \ln  \lambda = \varphi -  \frac{d \lambda}{\lambda } .
\end{eqnarray}
A  transformation between gauge equivalent gauges will be called a change of gauge, with  {\bf gauge factor} (or a rescaling by the {\bf scale transformation}) $\lambda$ and  {\bf gauge transformation} (\ref{gauge-transformation}).\\[0.5pt]  
c) An  equivalence class of gauges, denoted by $[g, \varphi]$, is called a  {\bf Weylian metric} on $M$ and a manifold with Weylian metric $(M, [g, \varphi])$  will be called a {\bf Weylian manifold}. \\[0.5pt]
d)   A gauge will be called {\bf semi-Riemannian}, if it is of the form $(g, 0)$, i.e., its length connection is zero. A Weylian manifold will be called semi-Riemannian, if it has a semi-Riemannian gauge.
\end{definition}

\begin{remark}{\em 
If one considers the trivial principal bundle  ${\cal S}M  \longrightarrow M$ with group  $(\R^+, \cdot)$ and total space  $M \times \R^+$, a connection is the same as a real-valued  1-form. Here the gauge group  $\Gamma (SM)$ is the set of (differentiable) maps $M \longrightarrow \R^+$. ${\cal S}M$ will be called the {\em scale bundle}. The group $\R^+$ operates on the connections as given in
 (\ref{gauge-transformation}). In order to define a Weylian metric, one has to specify a pair consisting of a connection in ${\cal S}M$ and a metric $g$ on $M$, and to introduce an equivalence relation as in b). Equivalently, if one starts from a conformal structure $[g]$ on $M$, represented by $g$, a Weylian metric is specified by the selection of a connection in the scale bundle. Change of representatives of $[g]$ go along with changes of the trivialization of ${\cal S}M$ and the corresponding gauge transformation for the 1-form. This view turns out to be useful when   conformal approaches to cosmology  or field theory are related to the Weyl geometric point of view.

Here we do not need the bundle structure and we work with our definition 1, which is introduced very much in the spirit of Weyl's original presentation.  It may be worthwhile to mention that the ``gauge transformations'' of our definition  are {\em gauge transformations} of connections in $SM$ in the sense of modern differential geometry.\footnote{After local trivialization, the connection can be expressed by a Lie algebra valued 1-form $A$; a change of gauge is expressed by the operation of a Lie group valued function $\gamma$. Because of commutativity the ensuing gauge transformation for the 1-form, $\tilde{A} = \gamma A \gamma ^{-1} + \gamma  d \gamma ^{-1}$, reduces here  to  $\tilde{\varphi} = \varphi+ \lambda   d (\lambda  ^{-1}) = \varphi - d \ln \lambda $. Of course, the sign is important for the consistency of gauge geometrical considerations, cf. footnotes \ref{fnDirac1}, \ref{fnDirac2}. \label{fn gauge transformation}}
  If other gauge structures enter the game, one ought to add the attribute ``scaling'', to make clear which (part of an perhaps extended) gauge group is referred to. In fact, the terminology ``gauge'' transformation is historically derived from this exemplar. 
}\end{remark}

In the sense of remark 1, the  curvature $F$ of a length connection is well defined. If the connection is given by the differential form $\varphi$, the curvature is  simple to calculate; it is nothing but the exterior differential
\beq \label{length-curvature} F = d \varphi .\eeq
$F$ is obviously gauge independent and will be called the {\em length curvature} or {\em scale transfer curvature} of the Weylian manifold.

For the sake of notational simplicity,  in the sequel  the symbol $g$ will often be used  as a shorthand expression for the local matrix   $(g_{ij})$ of $g$. 
Of course the expression of $g$ could  just as well be related to an orthonormal frame (ONF) description of the tangent bundle. That is   advisable in contexts  where physical reference frames are necessary in the theoretical construction, as for the description of 
the  general relativistic Dirac equation  \cite{Weyl:DiracZPh,Audretsch/Gaehler/Straumann} etc.. The geometry of  Weylian manifolds will be called Weylian {\em gauge geometry} (in the original sense of the word), or in short {\em Weyl geometry}.

\begin{remark} \label{isoWMf}
{\em 
An {\em isomorphism} of Weylian manifolds $(M, [g, \varphi ])$ and  $(\tilde{M}, [\tilde{g},\tilde{\varphi ]})$ is  given by a diffeomorphism $\beta: \; M \rightarrow \tilde{M} $, which transforms the Weylian metrics into each other, 
\[  \beta _{\ast} \tilde{g} = \lambda g, \ \beta _{\ast} \tilde{\varphi} = \varphi - d \ln \lambda , \]
where $\lambda $ is a gauge factor  and   $\beta _{\ast}$ is an abbreviated notation for  the linearization of $\beta $ operating on differential forms (of any degree). Clearly $\lambda $ depends on the choice of gauges in $M$ and $ \tilde{M}$. 
The defining data are $(\beta, [\lambda])$.
}
\end{remark}

According to Weyl's argumentation of 1918, any representative $g$ of the conformal metric may be used  for a direct metrical comparison of vector or tensor data at one and the same point $p \in M$,  or in its infinitesimal neighbourhood. The comparison of metrical data  given at  different  points $p$ and $q$ (``of finite distance'') can be achieved only indirectly \cite[etc.]{Weyl:InfGeo,Weyl:RZM5}. The length connection $\varphi$ can be used to ``transport length units''     along a (a sufficiently differentiable) curve $c(t)$.   More formall--y let us introduce: 
\begin{definition}
With respect to the gauge
 $(g, \varphi)$, a { \bf  length transfer}  or {\bf scale transfer} along a differentiable 
 curve  $c: [0, 1] \rightarrow M$  is defined  as the  real valued  function $l: [0,1] \longrightarrow \R^+$ with
\beq \label{form-length-transfer}  l(t):=   l(q) =  e^{ \int_0^t \varphi (c' (\tau)) d \tau }  . \eeq
 \end{definition}

\begin{remark} \label{remark length-transfer}
{\em   As $l$ is not dependent on the parametrization, we often  write $l(p):= l(c(t))$, if  a  curve $c$ is specified and $p=c(t)$.  A length transfer makes metrical quantities at $p = c(0)$ and $q = c(t)$ comparable, although its definition  depends on the gauge and the path. That is shown by the next lemma.  
One could equivalently define the length transfer by a differential equation
\[  l'(t)  = l(t) \varphi(c'(t)), \;\; l(0) = 1. \]
Weyl   preferred a differential point of view and expressed this condition in terms of the change of the ``infinitesimal length factor'' $dl$
\[ dl = l  \sum_i \varphi_i dx^i . \] 
Transformation under change of gauge is given by
\beq \label{l-tilde} \tilde{l} = e^{\int ( \varphi - d \log \lambda )} = e^{- \log \lambda} e^{\int \varphi} = \lambda ^{-1} l . \eeq
An immediate consequence is the following observation.  
} \end{remark}

\begin{lemma}
For vectors $\xi \in T_p M$ and $\eta \in T_q M$  and $l$  the length transfer with respect to  a  gauge $(g, \varphi)$, the modified quotient
\[  l(c(t))^2  \frac{g_{c(t)}(\eta,\eta)}{g_p(\xi, \xi)} \]
is  independent of the gauge. It  may be used for a gauge invariant comparison of lengths of vectors at different points of a Weylian manifold.
\end{lemma}

For a {\bf proof} one only needs to notice that the gauge transformations for the terms $l^2$ and $g$ cancel.\footnote{With Dirac's convention for ``gauge'' transformations (cf. footnote \ref{fnDirac1}) the gauge weight of length transfer changes sign and  the lemma no longer holds. \label{fnDirac2} }

 In general  the comparison
depends    on the choice of the path. For the case of integrable Weyl manifolds (see below), to which we
will restrict our view from the  following section onward, it does not.  
 In any case,  the length transfer can be used as a factor to modify the  measurement of vectors along a curve  in a given metrical component $g$ of a gauge $(g, \varphi)$; i.e., one measures the length by $ l^2 g$  rather  than  by $g$ and arrives at a gauge invariant concept of comparison.  

\begin{remark}{\em \label{length-transfer}
Weyl called this   comparison of vectors at  different points  ``calibration by transfer''. Its invariance under changes of gauge  was  behind his choice of terminology   ``gauge transformation'' in the literal sense of the word. 
If a curve $c$  with 
 $c(0)=p$ and $c(1)=q$ is given and  two equivalent gauges $(g,\varphi)$,  $(\tilde{g},\tilde{\varphi)}$,   with  length transfer factors  $l$ and $\tilde{l}$ respectively (along $c$), are related by a gauge transformation with gauge factor $\lambda$, $\tilde{g}= \lambda ^2 g$, the above equation
 (\ref{l-tilde}) is written more precisely  as
\[ \tilde{l}(p,q) = \frac{\lambda (p)}{\lambda (q)} l(p,q) .  \]
Thus length transfer functions are changed under a  gauge transformation by a gauge factor $\lambda (p)$ in the first variable  and by $\lambda ^{-1}(q)$  in the second  one. In  Weyl's terminology which will be explained in a moment (definition
 \ref{Weyl-field}),  length transfer is given by a function $l(p,q)$ of {\em gauge weight} $+1$ in the first variable and  $-1$ in the second.
}\end{remark}

  Weyl  gave strong arguments that his generalized metric results from an analysis of how one may understand   metrical relationships from a ``purely infinitesimal'' point of view and that it is a   better conceptual starting point than Riemannian geometry  for the development of physical geometry in the sense of general relativity. 

He considered it as a clue to his  gauge geometry that a Weylian manifold has a uniquely determined  compatible affine connection ({\em fundamental theorem} of Weyl geometry).   {\em Compatibility} means that parallel transfer of vectors preserves angles and that the length of parallel transported vectors changes with the  length transfer along the curve (which is gauge invariant by Weyl's lemma 1).  
\begin{theorem}[Weyl] \label{W-L-C}
A Weylian manifold $(M, [g, \varphi ])$ has a uniquely determined  affine connection $\Gamma $ compatible with the Weylian metric. If a gauge $(g, \varphi)$ is chosen, and  ${}_g{}\Gamma $ is  the Levi-Civita connection of  $g$ alone, the following relation holds:
\[  \Gamma ^i_{jk} =  {}_g{}\Gamma ^i _{jk} +  ( \delta ^i _j \varphi_ k   + \delta ^i_ k \varphi_ j  - g_{jk}  \varphi ^i )  \] 
(using Einstein sum convention for common coordinate indexes in the given quantities  and the Kronecker symbol $\delta $).
\end{theorem}
 
\noindent Clearly  $_g\Gamma $ is gauge dependent, while $\Gamma $ is not. For a {\bf proof} see, e.g., \cite[124f.]{Weyl:RZM5}. The theorem is behind the following
 
\begin{definition} The  uniquely determined compatible affine connection $\Gamma$ of a Weylian manifold $(M, [g, \varphi])$  will be called the {\bf Weyl-Levi-Civita} connection, or in short W-L-C connection, of $M$. By covariant derivation  in $M$ we refer to   the derivation $D = D_{\Gamma}$ with respect to the W-L-C connection, if not otherwise specified.
\end{definition}

As the Riemann curvature tensor of type (1.3), $R =(R^i_{jkl})$, can be geometrically defined from the affine connection only (without reference to the metric), the Riemann curvature of the W-L-C connection is independent of gauge. The same holds for   the Ricci curvature as  its contraction (which does not  involve the Riemannian component $g$ of the metric), whereas scalar curvature $\bar{R}$ as the contraction 
of  $R^i_j = g^{ik}R_{kj}$, with $(R_{jk}) = Ric$ is obviously gauge dependent (we will see in a moment that $\bar{R}$ is a scalar function of ``gauge weight'' $- 2$).  Let us resume this observation as
\begin{lemma}[Weyl]
The Riemann curvature of type (1,3 ),  $R =(R^i_{jkl})$, and the Ricci  curvature of the Weyl-Levi-Civita connection are well defined tensor fields (i.e. gauge independent) on a Weylian manifold.
\end{lemma}

It is therefore useful to agree upon the following terminology:
\begin{definition}
In a  Weylian manifold $(M,[g,\varphi])$ with Weyl-Levi-Civita connection $\Gamma$  we define:\\[0.5pt]
a) The {\bf Riemann} respectively {\bf Ricci } curvature of the Weylian metric are the respective curvature tensors $R$ and $Ric = (R_{ij})$ of $\Gamma$.\\[0.5pt]
b)  The {\bf scalar curvature} $\bar{R}$   is a family of scalar functions $\bar{R}_{(g, \varphi)}$  on $M$, indexed by the gauges $(g, \varphi)$, defined as  $\bar{R}_{(g, \varphi)}:= g^{ij}R_{ij}$ (using Einstein's sum convention).\\[0.5pt]
c) The {\bf length curvature} or {\bf scale transfer curvature   } $F$  is the (gauge independent) exterior differential $F = d \varphi$ for any gauge of $M$ (see equ. (\ref{length-curvature})).
\end{definition}

The integration of the infinitesimal length transfer (\ref{length-transfer}) is independent of the  path,
 iff the curvature of the length connection vanishes. That gives  an easy criterion   for the possibility to find a semi-Riemannian  gauge of a  Weylian manifold.

\begin{theorem}[Weyl]
 A  Weylian manifold $(M, [g, \varphi ] )$ is locally semi-Rie\-man\-nian,  iff its length curvature vanishes, \[F = d \varphi = 0.\] 
\end{theorem}

\noindent {\bf Proof} \quad In this case, a semi-Riemannian gauge of $M$  can be calculated from an arbitrary gauge $(g,\varphi)$  by
 the    path independent integrals starting from a  reference point $p$ (in any path connected component of $M$) and a path  $c: [0,1] \rightarrow M$ from $p$ to $q$ 
\beq \label{Riemann-gauge-factor}  \lambda (q) = e^{ \int_0^1 \varphi (c' (\tau )) d \tau } .  \eeq
As $d \varphi = 0$,  the length transfer function is path independent in simply connected regions and can be used as a gauge factor. This 
leads to a  $\tilde{g} = \lambda^2 g$ and $\tilde{\varphi} (q) = \varphi (q)  - d \ln \lambda (q) = \varphi (q)  - \varphi (q) = 0$ and thus to a semi-Riemannian gauge. Two semi-Riemannian gauges differ only by constants on each pathwise connected component of $M$.\\[0.5pt]

Vanishing of the length curvature is  an integrability condition and decides, whether a  Weylian manifold is semi-Riemannian. In the case it is, we will call the Weyl geometry    {\em integrable}. Although for integrable Weyl manifolds  it seems at first glance that  we can just as well go over to   the view  of semi-Riemannian geometry, we shall see that  Weyl geometry allows to consider physically non-trivial metrical modifications of semi-Riemannian geometry {\em even in this special case}.
Audretsch, G\"ahler and Straumann have shown that there are  strong reasons to restrict attention to integrable Weyl manifolds  in the physical context \cite{Audretsch/Gaehler/Straumann}. We will come back to their argument in the next section. Mathematically their argument builds on the following observation for Weyl manifolds:

\begin{lemma}[Audretsch/G\"ahler/Straumann]
\label{Audretsch_ea}
If  a Weylian manifold  $(M,$ $ [g,  \varphi])$ with W-L-C connection $\Gamma$ has a scalar Weyl field $f$ with vanishing covariant derivative,
\[  D_{\Gamma} f = 0 \; ,\]
 $M$ is integrable. 
\end{lemma}

\noindent {\bf Proof} \quad
If the scalar Weyl field is given by  $f$ in some gauge  $(g, \varphi)$, the covariant derivative is
\[   D_{\Gamma} f =  d f - f \varphi   \; , \] 
 where $d f$ denotes  the ordinary differential. Vanishing of the covariant derivative  implies 
\[  \varphi = \frac{d f}{f} =  d \log f \; .\]
Thus the length connection is a complete differential and integrable by the scale function $\lambda := f$.\\[1pt]

If a  tensor or a tensor field $X$ is defined mathematically or empirically in any way by some relationship to metrical quantities, it may change under a change of  gauge. Therefore we have to take into account  the gauge behaviour of vectors,  tensors, and their densities,  if we want  to do physics in the generalized geometrical framework  \cite{Weyl:InfGeo,Weyl:RZM5}.  
In order to give  a well defined  mathematical meaning to this concept,  we introduce: 
\begin{definition} \label{Weyl-field}
a) Let $(g, \varphi)$,  $(\tilde{g}, \tilde{\varphi})$ be gauges in a Weylian manifold $(M, [g, \varphi])$  and   $X, \tilde{X}$  tensor fields (or tensor densities) on $M$. We then  define triples to be 
{\bf k-equivalent},  
$  (g,\varphi, X) \sim _k (\tilde{g}, \tilde{\varphi}, \tilde{X}) $, iff  the gauges $(g, \varphi)$,  $(\tilde{g}, \tilde{\varphi})$ are gauge equivalent with  gauge factor $\lambda$, i.e. $ \tilde{g} = \lambda^2 g $, and 
\[\ \tilde{X} = \lambda ^k X  . \]
b)  A  ``gauge quantity'' or a {\bf Weyl field of gauge weight k}   is a $k-$equivalence class of  tripels  $ [g,  \varphi, X]_k$.
\end{definition}
 To simplify language, we often use representatives  $X$  of the third partner of the tripel and speak of tensors of weight $k$, if the context makes clear  which gauge $(g,\varphi)$  we refer to. Covariant differentiation of Weyl fields contain terms dependent on the gauge weight. We do not use it here and refer {\em in this respect} to \cite[sec. 3]{Dirac:1973}.

\begin{remark} \label{remark gauge-weight} {\em
For the sake of brevity, let us use double square brackets to indicate the gauge weight of geometrical or physical quantities, e.g. $[[g_{i j}]] = [[g]] = 2 $, as the metric field $X= g$ has  gauge weight 2 by definition; similarly  $[[g^{i j}]] =  -2 $.  We know already  that the Ricci tensor of a Weylian manifold is gauge invariant, i.e., its gauge weight is $ [[Ric ]] = 0$. The same holds, of course,  for all ``classical'' (gauge invariant) tensor (including vector and scalar) fields or densities.  
The scalar curvature $\bar{R} $ of a Weylian manifold   is  a scalar function of  gauge weight $[[\bar{R}]]= -2$, as can be  seen from the relationship $\bar{R} = g^{ij} R_{ij}$ ($Ric = R_{ij}$ in  coordinate notation).

It is obvious how to form  tensor products of Weyl fields of gauge weights $k$ and $l$. Clearly the product is a   Weyl field of weight $k + l$. Thus,  although neither the Riemannian component $g$ of a Weylian metric $[g, \varphi]$ nor its scalar curvature $\bar{R}$ are gauge invariant, their product is (gauge weight $ 2- 2 =0$). Therefore $\bar{R}g$ can be considered as   a ``classical'', i.e., gauge invariant, tensor field. 

 Time, the measure dimension of which will be denoted as $[t]$,  has gauge weight +1, as have lengths $[l]$, whereas energy as  its conjugate quantity (measure dimension $[E]$) is of   gauge weight  $[[E]] = -1$. 
Energy density is thus of gauge weight $[[E][ l^{-3}]]=  - 1 + (-3) = -4$ and so on.\footnote{Weyl considered the gauge factor of the metric itself as the main  reference for  weights. His gauge weights are therefore half of the ones used here. The latter are closer to common  dimensional considerations of physical quantities ( see, e.g.,  \cite{Manin:MathandPhys}).}

}
\end{remark}

The concept of length transfer can now be generalized to scalar Weyl fields of arbitrary gauge weight. Weyl introduced a suggestive terminology (``calibration by transfer'') which we want to adopt  in the following sense:
\begin{definition} \label{definition calibration-transfer}
If $l$   is the length transfer function along a curve $c:[0,1] \rightarrow M$ with respect to a gauge $(g, \varphi)$, a scalar Weyl field  $f$ of gauge weight $k$, defined along $c$, is said to  propagate by {\bf calibration transfer} along $c$, iff  for any pair of values $0\leq a \leq b \leq 1$  and  $c(a)=p, \, c(b)=q$ 
\beq \label{calibration-transfer}  f(q) = \left( \frac{l(q)}{l(p)}\right)^k f(p) . \eeq

\end{definition}

 An easy calculation shows that it is  possible to gauge a Weylian manifold with nowhere vanishing scalar curvature, $\bar{R} \neq 0$ everywhere,  such that the representative of scalar curvature becomes constant, e.g., $\bar{R} = \pm 1$. Starting from any gauge $(g, \varphi )$, this condition can easily be satisfied  by applying the gauge factor $\lambda^2 = |\bar{R}| $. This gauge transformation gives
\[ \tilde{g} = |\bar{R}| g =\lambda^2 g  \; , \; \; \tilde{\varphi} = \varphi - d \ln \sqrt{|\bar{R}|}  \]
and leads to $\tilde{R} =  \lambda^{-2} \bar{R} = \frac{\bar{R}}{|\bar{R}|} = \pm 1$, as wanted. 

\begin{lemma}[Weyl]
\label{Weyl-gauge}
For a Weylian manifold $(M, [g], \varphi)$ of nowhere vanishing scalar curvature $\bar{R}$, there is a gauge which    normalizes scalar curvature  to $\bar{R} = const$. Of course, such a gauge is unique up to a  constant. 
\end{lemma}
Weyl assigned this particular gauge a particularly important role in his  explorations of the use of Weyl geometry in cosmology, and we will follow and continue this track.  We therefore introduce the terminology: 
\begin{definition}
In a  Weylian manifold $(M, [g], \varphi)$ of nowhere vanishing scalar curvature, a  gauge with constant scalar curvature will be called a {\bf Weyl gauge} of $M$.
\end{definition}

\begin{example}{\em
Consider a {\em Robertson-Walker manifold} $M  = I \times_f S_{\kappa} $ with fibre $ S_{\kappa}$ over an open interval $I \subset \R$ with respect to a warp function 
$f: I \rightarrow \R ^+$, where $  S_{\kappa}$ is a 
 3-dimensional Riemannian manifold with metric $ d \sigma ^2 $
 of constant sectional curvature $ \kappa =  \frac{k}{a^2}$  ($ k = \pm 1$ or $0$ and $ a > 0 $) \cite{ONeill}.
 Using the parameter $\tau $ for elements in $I$, the Lorentzian warp metric $\tilde g $  on $M$ can be written as
\[ \tilde g: \;\;    ds^2 = d \tau ^2 - ( f d \sigma )^2 . \]
If we introduce the Weylian metric by the gauge $(\tilde g, \tilde{\varphi })$ with $\tilde{\varphi} =0$, we call $(M, [\tilde g, 0])$ a {\em Robertson-Walker-Weyl manifold}. We easily change the gauge to $(g_1, \varphi_1)$, such that one gets ``static rest spaces'', i.e.,  constant metric along the spatial fibres 
 \[  g_1 := f^{-2} \tilde g = \frac{d \tau ^2}{f^2} - d \sigma ^2  , \;\; \varphi_1 = - d \ln f^{-1} =   \frac{d f}{f} . \] 
One should {\em not be deceived} to think that now the spatial sections might be gauged to constant curvature in Weyl geometry. That cannot be the case, in general, as sectional and scalar curvatures are of gauge weight $ - 2$  and we know the values of these curvature parameters for Robertson-Walker manifolds in the semi-Riemannian gauge. The scalar curvature $\tilde{R}$ and sectional curvatures and  $\tilde{\kappa}_T$ and $\tilde{\kappa}_S$     in plane directions containing the vector $\frac{\partial}{\partial \tau}$  (``cosmic timelike'')  or   in plane directions tangential to the spatial fibres respectively  are in  semi-Riemannian gauge $(\tilde g , 0)$ (see e.g. \cite[345]{ONeill}):
\beqa \tilde{R} &=&  6 \left( (\frac{f'}{f})^2   + \frac{\kappa }{f^2} + \frac{f''}{f} \right) \, , \\
\tilde{\kappa} _T &=& \frac{f''}{f} , \;\;\;\;\;\;  \tilde{\kappa} _S = (\frac{f'}{f})^2 + \frac{\kappa }{f^2} \; .
\eeqa
We thus arrive at a {\em Weyl gauge} $(g, \varphi)$ by setting
\[  g:= |\tilde{R}| g , \;\;\; \varphi = - \frac{1}{2} d \ln |\tilde{R}| \, . \]
Now we have constant scalar curvature. To achieve constant sectional curvatures (in cosmic timelike {\em or} in spacelike direction) one has to use the respective values of sectional curvatures as  scale gauge factors. 
 }\end{example}

In the case of a  Weylian manifold of vanishing length curvature and nowhere vanishing scalar curvature, we  now have two distinguished gauges, {\em Riemann gauge} ($ \varphi = 0$) and {\em Weyl gauge} ($\bar{R} \equiv const$). 
 Even in the last edition of {\em Raum - Zeit - Materie}, two years after Weyl had given up his belief in the physical reality of his early  unified field theory,  he insisted upon the utility to study the interplay of both gauges. He  guessed that it might be particularly useful for the understanding of a hypothetical ``cosmological term'' in the (generalized) Einstein equation and might contribute to an understanding of cosmological redshift  \cite[300ff.]{Weyl:RZM5}.

\subsubsection*{2.  A conservative extension of  GRT }
Until about 1926/27, Weyl tended to bind the usage of his gauge geometry in physics to an interpretation   of the length connection $\varphi$ as the electromagnetic potential. Let us call this interpretation  the {\em e.m.p.-dogma} of gauge geometry. It was  not the only possibility, even at the time. In fact, Weyl attempted to explore the possibilities opened up by gauge geometry   for an understanding of cosmology,  in particular redshift,  at different occasions. These remarks  stood in tension, and sometimes even in sharp conceptual   dissonance, with the  e.m.p.-dogma.  By several reasons Weyl never  explored the possiblities of his gauge geometry  for cosmology in  a ``pure'' form, i.e., {\em separated} from the e.m.p.-dogma.  

Two scientific generations later, the dogma was broken. J. Ehlers, F. Pirani, and A. Schild  had a different perspective, when they used Weyl geometry for  a conceptual foundation of GRT (``constructive axiomatics'' in the terminology of  philosophers of science) \cite{Ehlers/Pirani/Schild}.
They showed that under very convincing simple assumptions for the propagation of light and for the free fall of test particles, which are minimal conditions for the validity of special relativity in the infinitesimal approximation, a conformal Lorentzian structure is specified (null geodesics as descriptors for the light ray structure). The trajectories of  freely falling test particles determine a projective class of affine connections (trajectories pass along the geodesics). If both sets of data are compatible (null geodesics of the light structure are a subclass of the affine geodesics), the structure of a Weylian manifold is obtained.
 Notwithstanding their foundational intentions, Ehlers e.a. left it open, whether Weyl's length curvature $F$ and the corresponding conserved current associated to the gauge symmetry $\R^+$  might  perhaps relate the gravitational field to another universally conserved current and may thus ``contain some physical truth'' \cite[83]{Ehlers/Pirani/Schild}. 
About that time, high energy physicists had come to the conclusion that deep elastic scattering of electrons at nucleons was approximatively invariant under mass/energy scaling, and Weyl's ``length'' gauge  arose new interest among physicists. Now the scale invariance and its breaking by a scale connection was reconsidered \cite{Deser:1970,Canuto:Scaling,Hehl_ea:Kiel_II}.

Audretsch, G\"ahler and Straumann damped too high expectations in this respect. They investigated Dirac and Klein-Gordon matter fields on a Weylian manifold $M$ (assuming that the topological obstruction for  existence of a spinor bundle over $M$ vanishes). In both cases a mass function $m$ on $M$ apppears in the equations, which acquires gauge weight $-1$ in the Weyl geometry setting. (In our terminology $m$ is a scalar Weyl field of weight -1.) They considered the series development of the solutions in rising powers of $\hbar$, the
 so-called WKB development,
\[   \psi = e^{\frac{i S}{\hbar}} \left(  \psi_0 + \frac{\hbar}{i} \psi_1 + \ldots  \right) \; , \] 
 investigated the 0-th order approximation $j_0 = (j_0^{\mu})$ of the current $j$ associated with the matter field $\psi$, respectively $\psi_0$. Following an idea of J. Audretsch, they proved an    interesting property:

\begin{proposition}[Audretsch/G\"ahler/Straumann] \label{Audretsch}
The 0-th approximation of the current $j_0$ in the WKB-development of a Dirac field or a  Klein-Gordon field $\psi$ with mass ``function''  $m$ (scalar Weyl field of weight -1) on a Weylian manifold with spin structure is geodesic, if and only if the covariant derivative $D$ of $m$  vanishes:
\[  D_{j_0} j_0 = 0 \longleftrightarrow D m = 0   \]
\end{proposition}

\noindent For a {\bf proof }, see the original publication \cite[47f.]{Audretsch/Gaehler/Straumann}.

Proposing  geodesity of the 0-th order of a Dirac or a Klein-Gordan current  as a compatibility criterion of Weyl geometry with quantum mechanics, Audretsch and coauthors drew the {\em striking conclusion that  only integrable  Weylian manifolds are compatible with quantum physics} (pay regard to  
their geometrical lemma (\ref{Audretsch_ea})). Moreover, the natural gauge condition $m \equiv const$ led them  to consider Riemann gauge as the preferred gauge in their context.

In consideration of the result of Audretsch, G\"ahler and Straumann, only integrable Weyl manifolds seem to  be  acceptable for our large scale context.\footnote{In the light of the scale invariance in deep elastic scattering experiments, it may be premature, however, to  accept the restriction to integrable Weyl geometry as definitive on all scales. The semi-classical equations used in the proposition apparently lose their meaning in the high energy/small scale domain and demand for ``second quantization'' or other formal tricks to draw information from them. }  
 We will develop   a modified version of what Weyl called an  ``extension of relativity'' inside  this restriction  \cite{Weyl:eGRT}.

We do not draw the consequence, however, that the 
gauge condition derived from the Dirac or Klein Gordan equations describes matter behaviour directly.   In this respect we keep closer to the view as it is developed in \cite{Canuto:Scaling,Tiwari:Weyl_geometry} and continue to respect 
 Weyl's warning that there may be reasons to assume that matter fields  behave as if they adapt to  local structure fields (``calibration by adjustment''). These  may lead to results different  from  ``calibration by transfer'' in the sense of gauge geometry. Although  we consider the ``short wave limiting'' argument of our authors  (as they call the 0-th approximation) as  compelling with respect to its structural constraints  (integrability of the length connection), we do not follow the   second step of their evaluation  (gauge condition $m \equiv const$). 

If one does not follow this step, it becomes possible to establish a {\em conservative extension}  of GRT, which allows for the  possibility that matter properties are represented by  a gauge condition  different from Riemann gauge, although, of course,  the standard assumption of Riemann gauge keeps its validity in most of the standard applications of GRT.  The specification ``conservative''  is here used just like in theory extension  in the sense of  the foundations of mathematics: In our case, the Einstein equation is lifted into an  extended physico-geometrical framework, but without any change of dynamical properties for the old models. On the other hand,  new models can be constructed which were not conceivable before. For the extended theory we sometimes use the abbreviatian  eGRT.

We start from considering the Einstein equation 
\beq \label{Einstein-equ} Ric - \frac{1}{2} \bar{R} g = 8 \pi \frac{G}{c^4} T  \eeq
under the aspect of scale gauge in Weyl geometry ($Ric$ the Ricci tensor  as above, $\bar{R}$  scalar curvature,  $T$ the doubly covariant energy-momentum tensor, $G$ the Newtonian gravitational constant, and $c$  velocity of light).

\begin{remark}{\em 
Let us start with an elementary heuristic observation. 
 The left hand side (l.h.s) of the equation is obviously gauge invariant (cf. remark
 \ref{remark gauge-weight}). For the r.h.s. we calculate gauge weights by the elementary assumption that the weights are consistent with the dimensional construction of the  quantities. We use simple square  brackets $[ \; ]$ to denote physical dimensions and double square brackets $[[\;  ]]$ for gauge weights of the quantities; $l$, $t$, $m$ $E$ are used as symbols for any length, time, mass, or energy quantity.  The Newton constant has dimension $[G] = [l]^3 [t]^{-2}[m]^{-1}  $. If we want to treat it consistently in a scale gauge geometric theory, it has to be represented by  a scalar Weyl field of   gauge weight 
\[ [[G]] = [[l]^3 [t]^{-2}[m]^{-1}] = 3 - 2 - (-1) = 2 \; .\]

The components of  the co-contravariant energy-momentum tensor ${T}^i_j$ is of the same gauge weight as   energy density,  $[[{T}^i_j ]] = -4$. As  we are working with a doubly contravariant version of the Einstein equation, which is exactly the one in which the l.h.s becomes gauge invariant,  the right hand side tensor is  $ {T}_{ij} = g_{i \mu }{T}^{\mu }_j$ and thus  of gauge weight $[[ {T}_{ij}]]  = -4 + 2 = -2$. Thus the whole r.h.s. is of gauge weight $2 -2 = 0$. It is composed of scale gauge covariant quantities in such a way, that as a whole it is gauge invariant (as it must be the case, if there shall be a chance for generalization of the equation to gauge geometry).

This  shows already by an   elementary observation that the Einstein equation ought to be generalizable to Weyl geometry without modification of the underlying dynamical law. 
}
\end{remark}
 
 In fact, it is easy to transfer the Hilbert-Einstein variational principle to the case of Weylian
 manifolds.\footnote{This holds even for the non-integrable case although, by the above mentioned reasons, we confine ourselves to the integrable case.}
We consider a Weylian manifold $M$ with metric $[g, \varphi ]$ with integrable length connection $\varphi$,  W-L-C connection $\Gamma$, curvature tensor $R$, Ricci curvature $Ric$, and scalar curvature $\bar{R}$. 
We suppose the  Langrange density of the gravitational field $\Gamma$ as  given by a gauge invariant density of the form of the Hilbert-Einstein action (equ. (\ref{Hilbert-Einstein-action}) below).
Because of the gauge weights $[[\bar{R}]]= -2$, $[[\sqrt{|g|}]]= \frac{ 1 }{2 }(2 \cdot 4)  = 4$, the representing object for the Newton gravitational ``constant'' has to be   a scalar Weyl field of weight $[[G]] = 2$ in order that (\ref{Hilbert-Einstein-action}) becomes gauge invariant. We arrive at  
\begin{theorem}
Let $M$ be an integrable Weyl manifold with  metric $[g, \varphi ]$, W-L-C connection $\Gamma$, scalar curvature $\bar{R}$, and   $G$  a 
 nowhere vanishing positive scalar Weyl field  of weight 2 on $M$.
 Then the  Hilbert-Einstein action of the pair  $(M, G)$, defined by 
\beq \label{Hilbert-Einstein-action} {\cal L}_{H} := \frac{1}{16 \pi  G} \bar{R} \sqrt{|g|}   \; ,\eeq
is scale-gauge invariant. If, moreover, a scale-gauge invariant Lagrange density of matter and 
non-gravitational fields $ {\cal L}_M := L_M  \sqrt{|g|}$ is given on $M$,
 the variational equation for the  combined action 
\[  {\cal S}_H + {\cal S}_M := \int    {\cal L}_H dx +  \int    {\cal L}_M dx \]
 is  the   Einstein equation  (\ref{Einstein-equ})  with an  energy momentum tensor  given  by
\beq \label{r.h.s.}  T_{\mu \upsilon  }  := - \frac{1}{2 \sqrt{|g|}} \frac{ \partial {\cal L}_M}{\partial g^{\mu \nu }} \; . \eeq
 \end{theorem}

\noindent {\bf Proof} \quad The variational calculation of the Hilbert-Einstein action can  be transferred from  semi-Riemannian   to integrable Weyl manifolds. We norm all the Weylian  metrics over which the variation takes place by the condition $G \equiv const$. That can be done like in the case of the Weyl gauge  (lemma (\ref{Weyl-gauge}); here one only has to  {\em divide}  by $G$, $\lambda^2 = G^{-1}$, if $G$ is not yet constant). Now the variational calculation proceeds formally like in the semi-Riemannian case:
\beqa
\delta  {\cal S}_H& =&  \int \frac{\partial {\cal L}_H}{\partial g^{\mu \nu }} \delta g^{\mu \nu } dx \\
\delta  {\cal S}_M& =& \int \frac{\partial {\cal L}_M}{\partial g^{\mu \nu }} \delta g^{\mu \nu } dx
\eeqa

Because of the gauge normation for $G$, we get
\beqa
\frac{\partial {\cal L}_H}{\partial g^{\mu \nu }}\delta g^{\mu \nu }  & =& \frac{1}{16 \pi G} 
\frac{\partial (\bar{R} \sqrt{|g|} )}{\partial g^{\mu \nu }} \delta g^{\mu \nu }  = \frac{1}{16 \pi G} 
\left(\frac{\partial  \bar{R}}{\partial g^{\mu \nu }} + \frac{\bar{R}}{\sqrt{|g|} } \frac{\partial \sqrt{|g|} }{\partial g^{\mu \nu }} \right)\sqrt{|g|}\delta g^{\mu \nu } \\
&=& \frac{1}{16 \pi G} (Ric - \frac{1}{2} \bar{R} g _{\mu \upsilon } ) \sqrt{|g|} \delta g^{\mu \nu } \; ,
\eeqa
just like in the classical case.\footnote{Compare, e.g., \cite[360]{Weinberg:Cosmology}. Here we use a different sign convention for the energy-momentum tensor.}
Note, however, that in this calculation the curvature expressions contain  contributions from the length connection $\varphi $, if  the chosen gauge  (called below ``gravitational gauge'') is not identical to Riemann gauge.
The condition of stationarity,
\[  \delta  {\cal S}_H + \delta  {\cal S}_M = 0 \; ,\] 
leads directly to the Einstein equation.
\\[1pt]

The gauge used above will obviously play  a preferred role in any attempt of establishing a reasonable semantics for Weyl geometric models in relativity. On the other hand,  for each empirical interpretation of a Weyl geometric model of general relativity we have to specify a  gauge  which coincides with the measurement of atomic clocks and distances measured by material devices.  
We therefore introduce
\begin{definition}
A Weyl geometric model of extended general relativity  (eGRT) is given by an integrable Weyl manifold together with an everywhere  positive  scalar Weyl field $G$ of weight 2. $G$ is called the ``gravitational constant''. A gauge for which $G$ is constant in the literal  sense  ($G(x) \equiv const$ for all $x \in M$) is called {\bf gravitational gauge} of the model. The gauge which expresses measurements by atomic clocks and material devices will be called {\bf  matter gauge}. 
\end{definition}

In order to avoid an unnecessary multiplicity of differing gauges, we will assume \[ \mbox{matter gauge} = \mbox{gravitational gauge} \]
 in the sequel.
Of course, this involves the assumption that    measurements by material objects lead to one and the same value of the gravitational constant anywhere and at any time in the physical ``cosmos''. If we wanted to analyze  the consequences of P.A.M.  Dirac's and P. Jordan's different assumption (of a time dependent gravitational constant) in our framework, we had to give up this identification. That  will not be done in this investigation, although the next example shows a weak reflection of Dirac's idea.. 

\begin{example}{\em 
Take $G := f^2$ in the semi-Riemannian gauge of a Robertson-Walker-Weyl manifold  (example 1). Rescaling by $\lambda  = f^{-1}$ gives the gravitational gauge  $(g_m, \varphi_m)$. Then the Riemannian component of the metric leads to   ``static rests spaces'',
 \[  g_m = \left(  \frac{d\tau ^2}{f} \right)  - d\sigma ^2 , \;\;\; \varphi_m = d \log f . \]
In our approach (not  in Dirac's and Jordan's !) the assumption of a ``changing gravitational constant'' {\em in semi-Riemannian gauge} leads to a physically preferred gauge  in which $G$ is unchanging. The latter is   different to the gauge  preferred by purely differential geometrical reasons. Moreover,  a reparametrization of the ``cosmic time'' parameter $t := \int \frac{d\tau }{f}$ simplifies the coordinate expression of the metric,
\[  g_m = dt^2 - d\sigma ^2  , \;\;\; \varphi_m = d \log f .  \]
Of course, $G$ is  a ``true constant'' in this gauge (we have chosen it so). 
}
\end{example}

Obviously the Weyl geometric extension of GRT is weak and conservative in the sense that any ``classical'' general relativistic model of semi-Riemannian geometry remains one in the extension, without any change of its geometrical properties or its  dynamics. It is, nonetheless, a veritable {\em extension}  (even though a much more modest one than originally intended by Weyl), because the integration of the Einstein equation allows here a change of scale gauge. Its solution acquires, so to speak, an additonal ``parameter of integration''. 

Because of the restriction to integrable Weyl manifolds, the scale factor and the corresponding scale connection $\varphi$ are constrained by the differential equation
\beq \label{length-curvature-0} d \varphi = 0 .
\eeq
The Einstein equation (\ref{Einstein-equ}) and the integrability constraint
  (\ref{length-curvature-0})  together characterize the physical geometry in our weakly extended general relativity (eGRT).

In this extension, the Einstein equation contains additional terms from the Weylian length connection, if  gravitational gauge (matter gauge) $(g, \varphi)$ is different from Riemann gauge, i.e. $\varphi \neq 0$. The Ricci tensor and scalar curvature of the Weyl-Levi-Civita connection $\Gamma$ can be decomposed into terms derived from the Riemannian component $g$ only and these additional terms. Let aus denote the Ricci tensor and scalar curvature of the Riemannian component $g$ of our gauge by $Ric_g$ and $\bar{R}_g$. Then the l.h.s of the Einstein equation is naturally decomposed into the classical Einstein tensor of the Riemannian component $g$ of the Weylian metric $(g, \varphi)$ and a term which contains all the contributions from the non-vanishing length connection. We denote the latter by  $\Lambda_{\varphi}$ and arrive at a decomposition of the form
\beq \label{Weyl-Lambda}  Ric -  \frac{1}{2}\bar{R} g =   Ric_g  -  \frac{1}{2}\bar{R}_g g + \Lambda_{\varphi} \; , \eeq
which can be read as a defining equation for $\Lambda_{\varphi}$. In our models this additional term plays a role analogous to the cosmological term of standard cosmology. Of course, we can only expect a  functional analogy.  $\Lambda_{\varphi}$ cannot be brought into the form constant times $g$. Moreover, it is gauge dependent and can be ``gauged away'' in Riemann gauge. The functional analogy is strong enough, however, to introduce the  following 
\begin{terminology}
The term 
$  \Lambda_{\varphi} $ of equation (\ref{Weyl-Lambda}) will be called the 
 cosmological term of Weyl geometry, or in short the  {\bf Weylian cosmological term} of the gauge invariant Einstein equation.
\end{terminology}
In fact, Weyl tried at several places to arrive at a more natural derivation of a cosmological term from his gauge geometrical approach. He tried to  subsume Einstein's cosmological term  as a ``special case'' of his  and to make Einstein's term look more ``natural''   \cite[46]{Weyl:eGRT}. This could not work out by different reasons.

We have already seen that our weak extension of GRT by lifting it to integrable Weyl geometry does {\em not} imply the introduction of a new dynamical field.
 We only gain the  liberty  to investigate models in which Riemann gauge and gravitational gauge (and with it  matter gauge) are no longer identical, like in the classical case.  So we have  new questions to ask for all classical cosmological models with long range extrapolation of geometrical and dynamical behaviour of ``the world'': Can we be sure that gravitational gauge and matter gauge are   identical with Riemann gauge, even   over long distances and into  regions of extreme curvature? Or can small modifications of gauge assumptions lead to simpler models in the mathematical or physical sense, although they may coincide with the  recieved models in many aspects (hopefully not for all) up to observational error? 
We shall see in the sequel, that such a tiny modification  pays out for  cosmological model building.

\subsubsection*{3. Cosmological redshift and Robertson-Walker-Weyl models}

If we want to think cosmology in geometric terms, we have to 
specify   a {\em cosmic time flow} $U$  and/or  an  {\em observer field} $X$ on the  manifold  $M$, i.e., an everywhere timelike future oriented vector field on $M$ of unit length. Each such  field allows to consider infinitesimal space-time splits, by considering $U^{\perp}$ (or $X^{\perp}$) at each $p\in M$ as the infinitesimal rest space at $p$ with respect to the class of (``cosmic'', or more general) observers. Usually one even assumes the existence of a global foliation of $M$, orthogonal to the field of time flow $U(p)$ at each point $p \in M$, with strong homogeneity and isotropy conditions on the spatial leaves. Moreover, observer and cosmic time flow fields are usually identified, $X = U$. That imposes  a natural restriction on the observer field  as a  {\em cosmic} or {\em comoving observer field}. 

All this translates to the gauge geometric context without major modifications. We only have to take into account that the  unit length condition results in gauging observer fields with weight $-1$. We have to consider
\beqa U \; \mbox{with respect to}\;\; (g, \varphi) \; \mbox{s. th.} \;\; |U(p)|_g = 1 , \\
 \tilde{U} \; \mbox{with respect to}\;\; (\tilde{g}, \tilde{\varphi}) \; \mbox{s. th.} \;\; |\tilde{U}(p)|_{\tilde{g}} = 1 .
\eeqa
Because of 
$ |\tilde{U}(p)|_{\tilde{g}} = \lambda |\tilde{U}(p)|_{g}  $,  this implies
\[ \tilde{U}(p) = \lambda ^{-1}U(p) \, . \]

For a consistent use of gauge geometry in cosmology, a {\em cosmic time flow field $U$ or an observer field $X$} has thus to be represented by a   timelike future oriented  Weyl vector field $[X]$ of gauge weight $-1$  with unit length in every gauge. 

\begin{example}{\em 
Take  a 
Robertson-Walker-Weyl manifold $(M, [\tilde{g}, 0])$ as in example 1, with $\tilde{g} = d\tau^2 - (f d \sigma )^2$,  and define the observer field  in Riemann gauge  $(\tilde{g},0)$ as given  by
 \[ \tilde{X} (\tau_0) := \frac{\partial}{\partial \tau}_{|\tau_0} .\]
If  Riemann gauge is considered as the  matter gauge of the model, we arrive at standard cosmology described in the Weyl setting. The choice of any other gauge as matter gauge  will lead to new models in Weyl geometry.
}\end{example}

\begin{terminology} a)  A scale integrable Weylian manifold with the specification of a matter gauge $(g_m, \varphi_m)$ (equivalently gravitational gauge, i.e., $G \equiv const$) and a  time flow (comoving observer) field  $X$ of gauge weight -1 (future oriented time-like and unit length)   is called a {\bf  time flow model} of eGRT. The defining data are $(M, (g_m, \varphi_m), X)$. \\[0.5pt]
b) We will call a model like in example (3) a {\bf Robertson-Walker-Weyl model}.  It will be called {\em standard}, iff the reference (matter) gauge is the Riemann gauge of the Weylian manifold. 
\end{terminology}

\begin{remark}{\em \label{iso-time-flow-model} 
An {\em isomorphism between two time-flow models} $(M,  (g_m, \varphi_m), X)$ and $(\tilde{M}, (\tilde{g}_m, \tilde{\varphi}_m), \tilde{X})$ is given by an  isomorphism of the Weylian manifolds   $(\beta , [\lambda ])$  (as in remark (\ref{isoWMf}) and   a point-dependent  family of Lorentz transformations $\alpha = \alpha  (p)$  in  the tangent fibres of $M$, such that
\[   \tilde{X} (\beta (p)) = \beta ^{\ast}_{|p} \alpha (p)(\lambda (p) X(p)) \] 
(where $\beta ^{\ast}$  denotes the operation on tangent vectors etc. induced by the differential of $\beta $). Such an isomorphism allows to (Lorentz-) ``rotate'' the observer field at any point, in other words to transform between different observer fields by ``local'' Lorentz transformations, besides the application of an isomorphism of Weylian manifolds.
}
\end{remark}

For our investigation, the   expression of  cosmological redshift by the scale connection of matter gauge will be important. 
The motion of photons is described by null geodesics, the energy transfer of a photon (respectively the redshift of its frequency) between a point  $p$ of emission  and a point $q$ of observation is measured with respect to the local space-time split defined by an observer flow $X$ on $M$ (assuming that the emitting oscillator moves with $X$). The oscillation time $T$ of the photon, its frequency $\nu $ and its energy $E$ are related by:
\[   E = \hbar \nu  \; , \;\;\; E T = \hbar . \]

 Redshift is calulated by
\[ z(p, q)  = \frac{T'}{T} -1 =  \frac{\nu}{\nu'}. \]

In classical GRT the 
transfer of photon oscillation time from an emitting system kinematically  described by $X(p)$ to an  observer described by $X(q)$, respectively of photon energy,  is mathematically represented  by the orthogonal projection of the tangent of a  nullgeodesic on the  timelike directions of the comoving observer flow  $X(x)$ at $p$, respectively at $q$. Parallel transport of  tangents along null geodesics is well defined in Weyl geometry  by the Weyl-Levi-Civita connection. One might be tempted  to calculate the redshift $z$ of a photon passing from $p$ to $q$ in a Weylian manifold along a null geodesic $\gamma (t), \; 0 \leq t \leq 1$, with respect to the observer flow $X$  in eGRT like in GRT. Comparison of quantities derived from  the orthogonal projections on a  vector field $Y$ on $M$ by  $g_{(\gamma (t))}(\gamma '(t), Y(\gamma (t)))$ would not make sense in gauge geometry, however, even for a gauge invariant field $Y$.  One  has to take into account the calibration by transfer of the metric $g$ (definition (\ref{definition calibration-transfer})), and compare the recalibrated quantities  $l^2(t) g_{(\gamma (t))}(\gamma '(t), Y(\gamma (t)))$. This is a gauge invariant expression (the gauge weights of $l^2$ and $g$ cancel, as the weight of length transfer in dependence of the endpoint is $[[l]]=-1$, cf. remark (\ref{length-transfer})). 

As the  observer (time-flow) field $X$ is of weight $-1$, it is   natural  to give a gauge invariant characterization  of redshift in Weyl geometry, which  coincides with the classical one in Riemann gauge,  in the following way.:

\begin{definition}  \label{definition redshift} 
For a  time-flow Weyl  model $(M, (g_m, \varphi_m), X)$, let $\alpha (t)$  be defined along any nullgeodesic  $\gamma : [0  ,1 ] \rightarrow M$ with  $\gamma (0) = p$ and $\gamma (1) = q$ by  orthogonal projection on the observer field,  corrected by  length transfer,
\beq  \label{projection} 
\alpha (t) := l^2(t) g_{\gamma (t)}(\gamma '(t), X (\gamma (t))) \, ,
\eeq
where $l(t): =  l(\gamma (t)) $ is the length transfer function along the geodesic, normed by the condition $l(0):= 1$.
 
The {\bf redshift function} $z$ of the model is then defined on the subset $Z \subset M \times M$ of nullgeodesically connected pairs of points $(p,q) \in M \times M$ by
\beq  \label{redshift-function} 
   z(p,q)+1 := l(1)   \frac{\alpha (0)}{\alpha (1)}  \,  .   \eeq
\end{definition} 
\noindent With these specifications we get:

\begin{lemma} \label{gauge invariance of redshift}
The  redshift function $ z(p,q)$ in a time-flow Weyl model is gauge invariant. Its value  is identical to the redshift calculated in the semi-Rie\-mann\-ian gauge  according to the principles of standard cosmology.
\end{lemma}

\noindent {\bf Proof} \quad  The gauge invariance is a consequence of cancelling of the gauge weights of the factors:
\beqa  [[z + 1 ]] &=& [[l]] - [[\alpha ]]  =  [[l]] - ( 2 [[l]] + [[g]] + [[X]]) \\
&=&  -  [[l]] - [[g]] - [[X]]  = + 1 - 2 + 1 = 0 
\eeqa
In semi-Riemannian gauge the length transfer is constant, $l \equiv 1$; thus equations (\ref{projection}), (\ref{redshift-function}) reduce to the standard formula in classical relativity.\\[1pt]

We have thus arrived at a conservative implementation of redshift in eGRT. 
 Looking at different gauges of  physical geometry is the symbolical equivalent of admitting that we do not have direct knowledge about ``real'' expansion of spatial sections. In fact, such a property is  never directly  observable and depends  on the specific gauge condition imposed (knowingly or unknowingly). In this sense, it is  a theoretic construct. Trying to find out different gauges and what they allow to make understandable, corresponds to {\em adapting our geometrical frame to matter and energy structures the best we can}. It may turn out, that expansion fits best, but it may be that different answers will stand at the end.

\begin{remark} \label{remark wave-vector} {\em 
In standard relativity, GRT, the wave vector $k(s)$ of a plane wave modelling a photon semiclassically is bound to  a null geodesic $\gamma (s)$ by the  relation 
\[  k (s) = C \, \gamma '(s)  \]
(where $C$ is a constant to norm the  ``emission energy'' $E(s_0)$.) 
Energy decrease (redshift) of a photon has to be expressed by the relation of $\gamma '$ to the observer field. In classical relativity any kind of  energy decrease of photons, including gravitational one (like, e.g., in the Schwarzschild solution), has in this sense a ``quasi-kinematical'' formal characterization. Although this method offers,  for most cases, a well developed and consistent possibility to represent photon energy in differential geometric terms, it is not the only one.  In  Weyl geometry, it is natural to   renorm     the  wave vector by the length transfer function $l(s)$ along null geodesics,
\beq \label{wave-vector} k(s) = C \, l(s) \gamma '(s)  .\eeq
 It     is even compelling to do so, if one wants to uphold consistency under gauge change.  $k(s)$ then becomes  a gauge quantity (a Weyl field). As geodesics are defined gauge invariantly, the gauge weight of a wave vector field $k$ along $\gamma $ is clearly $[[k]] = [[l]] = -1 $, which is consistent with the gauge weight of energy. The  energy  of a photon for an observer with time flow field $X$  develops  along $\gamma $ according to
\beq \label{photon-energy} E(p) = g_p(k(s), X(p)) = g_p(l \gamma ', X) \eeq
(setting here $C=1$).
That is consistent with equations (\ref{projection}) and (\ref{redshift-function}),
\[ E(p) = l g_p(\gamma ' , X) =   \frac{\alpha (s)}{l(s)} = (z+1)^{-1} (\gamma (s))  . \]

In the expression $E = \alpha (s) l^{-1}(s)$, the first factor  $\alpha (t)$ may  be considered as a ``kinematical contribution'' to  redshift. It depends  essentially  on the choice of the observer field  and includes space kinematical effects. The second factor,   $l^{-1}(t)$, characterizes something like a  ``gravitational'' or, more general, ``tired light contribution'' to redshift (imputed by the gauge). The quotation marks  indicate here that,  in  general,  this distinction  is only of formal value. A change of gauge transforms the components into one another, as demanded by the equivalence principle.
In a gauge in which the ``kinematical contribution''  vanishes, $\alpha \equiv 1$, the  cosmological redshift appears as  completely due to the energy damping expressed by the length transfer $l(t)$. The above equation shows that in this case the photon energy propagates by calibration transfer of  weight -1,  in the sense of Weyl geometry (definition \ref{calibration-transfer}). 

This remark may seem abstract and methodological. The investigation of Robertson-Walker-Weyl models   shows, however, that we can really calculate with the effects of equivalent exchange between ``kinematical'' and ``gravitational/tired light'' contributions to cosmological redshift by an appropriate  change of gauge. 
Used in this way, {\em Weyl geometry allows to give an enhanced mathematical expression to the equivalence principle}.
}
\end{remark}

\begin{example}{\em 
In a Robertson-Walker-Weyl model 
\[  M = I\times_f S_{\kappa} , \;\; \tilde{g}: d \tau^2 - (f d \sigma )^2, \;\; \tilde{X} = \frac{\partial}{\partial \tau} \]
with {\em arbitrary} matter gauge (i.e., independent of whether we assume the model to be standard or not), 
the redshift of a photon  emitted at a point $p = (\tau, x)$ and observed at $p_0 = (\tau_0, x_0)$ (with $\tau < \tau_0$ and $x, x_0$ such that $p$ and $p_0$ are nullgeodetically connected) is given by
\[ z(p,p_0) + 1 =  \frac{f(\tau_0)}{f(\tau)} .\] 
This follows from standard results on classical Robertson-Walker models and our lemma (\ref{gauge invariance of redshift}).

For Robertson-Walker models, cosmological redshift depends only on the cosmological ``time'' parameters of the points of emission and observation. Thus there is   a natural extension of  the redshift function to all of $M \times M$, given by the r.h.s of the above formula.
Keeping the observer $p_0$ fixed and varying $p$, we consider $z(p,p_0)$ as a function of $p$ only, $z(p):= z(p,p_0)$. It even makes sense to  write  $z(\tau)= z(p_{\tau})$ in short, for any $p_{\tau }$ with time coordinate $\tau $. We now look for a gauge in which the length transfer function satisfies
\[  l(\tau) = z(\tau)+1 = \frac{ f(\tau _0) }{f(\tau ) } \, .\]
This  is achieved by applying the  scale gauge factor
\[  \lambda (\tau ) = \frac{ f(\tau _0) }{f(\tau ) } \]
to the semi-Riemannian gauge. Then 
\[  g  := \lambda ^2 \tilde{g} =    f ^2 (\tau_0) \left( (\frac{d \tau}{f} )^2 - d \sigma ^2  \right) \, , \; \; 
\varphi = - d \ln \lambda  = f(\tau _0) d \ln f\]
satisfies the condition. 
Note that in this gauge, which will be called {\em Hubble gauge} in a moment, the warping
of the fibre metrics is exactly compensated by the rescaling factor $\lambda (\tau ) = \frac{ f(\tau _0) }{f(\tau ) } $. We have thus arrived, in this gauge, at an unwarped static Riemannian component of the Weylian metric like in example 2.
In general, the infinitesimal linearization of redshift with timelike distance 
\beq \label{Hubble factor}\varphi (X) = (f' dt) \frac{\partial }{\partial t} = f' \eeq
is not constant.
}\end{example}

 E. Hubble was always careful not to claim that he had detected an ``expansion'' of the observed universe, but only a systematic relationship between redshift and distance. In the gauge above, the redshift is completely encoded in the Weylian length connection $\varphi$, independent of the answer to the question, which gauge is ``physical'', i.e., the matter gauge. 
Of course, Hubble  left it open to  more precise measurements in the future, whether the linearity of the dependence could be extrapolated. In any case, the latter is an interesting theoretical possibility.
These are  reasons enough to introduce the following
\begin{definition} \label{cosmological-model}
a) If  a time-flow Weyl model $(M, (g_m,\varphi_m), X)$ has  a gauge $(g, \varphi)$, such that the redshift $z(\gamma (t))$ along any nullgeodesic $\gamma $ is determined by the length transfer $l(\gamma (t))$ only  (i.e., $\alpha = l^2 g(\gamma ',X)= const$),
\beq \label{Hubble-z} z(\gamma (t)) + 1 = l (\gamma (t))  ,\eeq
we call $(g, \varphi)$ the {\bf Hubble gauge} of the model and $\varphi$  its {\bf Hubble form}.\\[1.5pt] \noindent
b) If for a Hubble gauge of a time-flow model the linearized redshift-time dependence is  constant,
\[  \varphi (X)   = const =: H \; , \; \; H \in \R^+ \cup \{ 0 \} \; , \]
the Hubble form will be called {\bf time-homogeneous} and $H$   its {\bf Hubble constant}.
\end{definition}

In the sequel we will use the terminology {\em cosmological Weyl models} for those time-flow models  which possess a Hubble gauge. This is the case for all R-W-W models (last example).
Obviously   ``the'' Hubble gauge is determined only up to a (global) constant factor, if it exists. Moroever, it is clear  that time-homogeneity results in a constant factor for the linearized redshift-distance relationship {\em in Hubble gauge only}. 

We have now identified three gauges in Robertson-Walker-Weyl models, which are of  theoretical interest: Riemann gauge, Weyl gauge, and Hubble gauge. The first two are always different, if warping is not trivial ($f  \not\equiv  const$). The last two differ in general, but may coincide in special cases. In the following we establish criteria for their   coincidence. 

\begin{proposition}
 In a Robertson-Walker-Weyl cosmology 
\[  M = I\times_f S_{\kappa} , \;\; \tilde{g}: d \tau^2 - (f d \sigma )^2, \;\; \tilde{\varphi} = 0, \;\; \tilde{X} = \frac{\partial}{\partial \tau} , \]
with any matter gauge the following holds:\\[1.5pt] \noindent
a) Hubble gauge and Weyl gauge coincide, iff
\beq \label{Hubble-Weyl-equ} f'' f + f'^2 = const . \eeq
b)  In particular Weyl gauge and Hubble gauge are identical, if the Hubble gauge is time homogeneous.
\end{proposition} 

\noindent {\bf Proof}  \qquad  Let us call $\lambda _W$ and $\lambda _H $ the gauge functions from Riemann gauge to Weyl gauge and  Hubble gauge respectively (see examples (1) and (4)),
\beqa 
 \lambda _W^2 &=& |\tilde{\bar{R}}| =  6 \left| (\frac{f'}{f})^2   + \frac{\kappa }{f^2} + \frac{f''}{f} \right| ,\\
\lambda _H &=& z(p) + 1 = \frac{const}{f(\tau)} \, .
\eeqa 
a) We thus find $\lambda _H \sim  \lambda _W$, iff
\[  \left| (\frac{f'}{f})^2   + \frac{\kappa }{f^2} + \frac{f''}{f} \right| \sim  \frac{1}{f^2} ,  \]
Multiplication by $f^2$ leads to  (\ref{Hubble-Weyl-equ}). The argumentation is  invertible.\\[1pt] 
\noindent
b) is an immediate result of the definition of time-homogeneity, equation 
(\ref{Hubble factor}), and a). \\[1pt]

\begin{remark}{\em 
 Hubble  and Zwicky suspected that the physical reason for cosmological redshift is not to be looked for in expansion of space. If this is correct,  redshift is much  better modelled by a Hubble form in the extended relativistic framework of integrable Weyl geometry than by a formal ``expansion factor'', even if reinterpreted by the equivalence principle. In such a  case,  Hubble gauge has to be accepted as the preferred ``physical gauge'' of the model, i.e., its matter gauge, rather than the traditional default choice of the Riemann gauge.
It is very interesting to notice that the simple assumption of time-homogeneity of 
Robertson-Walker-Weyl models (which in the  cosmological discourse about the middle of the last century used to be called,  slightly metaphysically, the ``perfect cosmological principle'') leads to the consequence that {\em  Hubble gauge and Weyl gauge are identical}. 

If the Zwicky-Hubble assumption is right, we are able to describe the physical geometry of the cosmos by a ``Weylianized'' Robertson-Walker  type structure. If, moreover, the laws of physics (including the reasons for the cosmic tiring of photons) are time-homogeneous,  material bodies behave as if they were gauged according to  scalar curvature. This need not be considered as a ``real'' process of  physical calibration, but can be understood as a geometrical  consequence of  the physicality of the Hubble form and of time-homogeneity. The assumption of  a ``real physical calibration'' of material bodies to the scalar curvature 
 was introduced as an interesting, but  highly ad-hoc argument by Hermann Weyl during the discussion of the question whether  his gauge geometry might be useful for physics,  in spite of A.  Einstein's criticism. He came back to this idea in his attempts to apply gauge geometry to cosmological modelling in the fifth edition of his book on {\em Raum - Zeit - Materie} \cite[299f]{Weyl:RZM5}.

 Here we have arrived at the {\em same formal result} from  different basic assumptions. 
}
\end{remark}

 It  thus seems legitimate  to call a  spatially  maximally homogeneous (and therefore also isotropic) Weyl model which possesses a time-homogeneous Hubble form a  ``Weyl universe''.  In the following definition,  we take  A. Walker's results into account, which show  that, in the semi-Rie\-mann\-ian view,  such models can be described by Robertson-Walker manifolds. Moreover we use the simplification of the metric by the Hubble gauge as shown in the last example.   

\begin{definition}  \label{Weyl-universe}
A cosmological Weyl model  ${\cal{M}} := (M,  (g,\varphi), X)$ with matter gauge $(g, \varphi)$ and observer field $X$ in this gauge will be called a {\bf Weyl universe}, iff the following holds: \\[0.5pt]
\noindent
 (i) $M$ can be written as a product  manifold  
\[M = \R \times S_{\kappa} , \]
where
 $S_{\kappa}$  is a Riemannian 3-manifold with   3-metric    $d\sigma ^2 $ of constant (ex-ante) sectional curvature $\kappa = \frac{k}{a^2}$  ($k \in {0, \pm 1}, \; a \in  \R^+$), \\[0.5pt]
\noindent
(ii) the Riemannian component of the gauge $(g, \varphi)$  is  given (in geometrical units, $c = 1$,  and using coordinates $ (x_0, x) \in \R \times S_{\kappa}$)  by
\beq g: ds^2 =   dx_0^2 - d\sigma ^2 ,  \eeq
(iii)  the corresponding scale connection is of the simplest non-trivial type
\beq  \varphi = H d x_0 , \; \;\;  H \in \R^+ \cup \{ 0 \} \; ,\eeq
(iv) and the observer field is 
\[   X:= \frac{ \partial  }{\partial x_0 }\, .  \]
\end{definition}

In physical units, the Weylian metric contains the velocity of light $c$, with $c  dt = d x_0$, and the Hubble connections acquires the form
\[\varphi = H d x_0 =  c H dt = H_0 dt\; ,  \]
where $ H_0 := c H$ is
the usual Hubble constant of astronomers (while our $H$ is often written as $H_1 = c^{-1} H_0$). On $S_{\kappa }$ we  prefer to use  Riemann's coordinates $(x_1, x_2, x_3)$, because  then the metric acquires the well known  homogeneous form  on spatial sections and shows  the close approximation to Euclidean space:

\[  d \sigma ^2 = \frac{ \sum \limits_{\alpha =1}^{3} d x_{\alpha }^2}{( 1 + \frac{\kappa }{4 }\sum \limits_{1}^{3}  x_{\alpha }^2)^2 } \; .\]

\subsubsection*{4. Basic properties of Weyl universes }
We know from the definition of Hubble gauge and from proposition 1 that the defining gauge $(g, \varphi)$ of a Weyl universe is  its  Weyl gauge. 
Because of integrability, a Weyl universe possesses a Riemann gauge. The scaling function $\lambda$ from Weyl gauge to Riemann gauge is calculated by integrating along ``time-flow'' lines $x_0 \in [a,y]$ lifted from the first component $\R$ to $M$,
\[ \lambda_{a} (y)  = e^{\int _a^{y} \varphi (\gamma '(\tau )) d \tau } = e^{H (y- a)} \, .\]
Along curves in spatial sections, $\lambda$ is constant, therefore the notation $\lambda_a(y)$ is justified. Usually we norm $a:= 0$; then we get
\[ \lambda (y) := \lambda _0 (y) = e^{H y} ,\]
or in physical units
\beq  \label{lambda-Weyl-universe} \lambda (t) = e^{c H t} = e^{H_0 t}  . \eeq
Riemann gauge  becomes $\tilde{g} := \lambda ^2 g$ and is given by the quadratic form
\[  d\tilde{s} =  e^{2 H x_0} d x_0 ^2 -   e^{2 H x_0} d \sigma ^2 .\]
Reparametrization by
\beq  \label{tau von x0} \tau = H^{-1} e^{H x_0} \eeq
 (like in example 2)  leads to 
\beq \label{Rob Walk picture} d \tilde{s}^2 = d \tau ^2 - (H \tau )^2 d\sigma ^2 \eeq
(of course, here  $\tilde{\varphi}=0$). The observer field is gauged into
\[  \tilde{X} = \lambda ^{-1}X = e^{- H x_0} \partial x_0 = \partial \tau \]
(using the abbreviation $\partial x := \frac{\partial }{\partial x }$) and is represented in the new  coordinate $\tau $ by the same coordinate vector field $(1, 0,0,0)$ as $X$ with respect to the $x_0$ coordinate. We  arrive at 
\begin{lemma}
A Weyl universe with Hubble constant $H$ is  a Robertson-Walker-Weyl model with linear warp function 
\[  f(\tau ) = H \tau ,\]
which has Weyl gauge as matter gauge.
\end{lemma}
In semi-Riemannian gauge, Weyl universes therefore 
look like  the simplest cases of (linearly) ``expanding'' universes, while   in Weyl gauge the Riemannian component of the Weyl metric is a static Lorentz type product of $\R$ by $S_{\kappa }$. According to the  different types of sectional curvature we introduce: 
\begin{terminology}{\em 
a) For  sectional curvatures $\kappa= \frac{k}{a^2 }$, $ k = \pm 1, 0$ we speak of
{\em Einstein-Weyl universes} in the case $k = +1$, and of {\em Minkowski-Weyl} or {\em Lobachevsky-Weyl universe} in the cases $k=0$ or $k=-1$ respectively.\\[1pt]\noindent
b) The  {\em module} $\zeta $ of a Weyl universe with $H > 0$ is defined by
\beq \label{zeta}  \zeta := \frac{\kappa}{H^2 } = \frac{k}{(aH)^2 }\; ;\eeq
$a$ is called the {\em ex-ante radius of curvature} ($k \neq 0$).
}
\end{terminology}

The terminology ``module'' is chosen because of
\begin{proposition}
a) Two Weyl universes with modules $\zeta $  and $\tilde{\zeta }$ are {\em isomorphic}, iff
\[ \zeta  = \tilde{\zeta }\, . \]
b) For $H=0$ there are three isomorphy types, characterized by $k$. 
\end{proposition}

{\bf Proof} \quad a) Consider first the case $k \neq 0$:\\
 ``$\Leftarrow$:'' $S_{k/a^2}$ is transformed by a homothety with factor $\eta := \frac{\tilde{a}}{a}$ into $S_{\tilde{k}/\tilde{a}^2}$. Thus we apply  a constant gauge factor $\lambda (x) \equiv \eta $. The length connection remains changed, $\tilde{\varphi} = \varphi - d \ln \lambda = \varphi = H dx_0$. We rescale the observer field by $\lambda ^{-1} = \eta ^{-1}$,
\[  \tilde{X} = \partial \tilde{x_0} =  \eta ^{-1} {\partial x_0},\]
which implies $ d \tilde{x_0} = \eta d x_0$ and 
\[\tilde{\varphi} = H dx_0 =  \eta ^{-1} H d \tilde{x_0}. \]
We thus arrive at
\[ \tilde{H} = \eta ^{-1} H = \frac{a}{\tilde{a }} H,  \]
and therefore $\zeta  = \tilde{\zeta }$.  

``$\Rightarrow$:'' If, on the other hand $aH \neq \tilde{a}\tilde{H}$, a homothety necessary to bring the metrics to coincide in spatial directions, does not transform the length connections into each other.
 \\[0.5pt]\noindent
 In the case $k=0$ and any  $H > 0$  we rescale by $\lambda = H$. This  leads to $\tilde{H} = \tilde{H}^{-1} H = 1$; thus all non-trivial k Minkowski-Weyl universes are isomorphic.  \\[0.5pt]\noindent
b) For $k=0, H=0$, we have the  Minkoswki space with trivial length connection, $\varphi = 0$. It  cannot be transformed  isometrically into a form with $H\neq0$. In this sense, it is degenerate. Similarly for  the other cases with $H=0, k=\pm 1$, the classical Einstein universe and its hyperbolic relative, the static ``Lobachevsly universe''.\\[1pt]\noindent

\begin{remark}\label{Milne} {\em 
The non-degenerate Weyl universes ($H \neq 0$) form a continuous 1-para\-me\-ter space of isomorphy classes, characterized by the module $\zeta $. The Hubble constant $H$ is a (global) scaling quantity. 
Empirically it has  been determined as
$H = H_1 = c^{-1} H_0 \approx 2.3 \pm 0.2 \, (10^4 \, Mpc)^{-1}$.\footnote{Corresponding to the value $ H_0 \approx 70 \pm 7 \, kms^{-1} Mpc^{-1}\approx 1.4 \pm 0.1 (10^{10}\, Y)^{-1}$. } 
For historically minded readers it may be worthwhile to observe that Milne's ``kinematical cosmology'' can now be re-read as being ``nearly'' a Minkowski-Weyl universe with $\zeta =0$ (a linearly expanding Minkowski space in the Robertson-Walker picture). The whole model class can be seen as a kind of  $\zeta $-deformation of this generic object. It is a pity that  Milne had no idea of the Weylian metric as a unifying concept;\footnote{Milne could have had one; all mathematical tools were present at the time. That Weyl universes have not been considered much earlier, seems to be a good case for a ``post-mature development'' in the sense of J. Stachel \cite{Stachel:Postmature}.}
 he rather used ``two metrics'' (corresponding to the Riemannian components of our Weyl  respectively Riemann  gauges) and preferred the ``wrong one'' from our point of view. He was fond of the expansionary picture, not only by physical reasons \cite[61, 66]{Kragh:Cosmos}.

For  empirical studies of cosmology in  the framework of Weyl universes, it will be crucial to determine an observational value for $\zeta$ on the basis of the most reliable  data we have and to investigate, whether the result is consistent with the evidence of  diverse other classes of observations. It is interesting to see that the supernovae luminosity data of the {\em Cerro Tololo Inter-American Observatory} group (CTO)   and  the {\em Supernova Cosmology Project} (SCP)  \cite{Perlmutter:SNIa} are 
as well represented by  Weyl universes as in the
 Friedman-Lema\^itre model class, although in the latter  a rather unnatural second parameter, the cosmological constant, has been introduced into the model space to enhance flexibility.\footnote{See the beautiful discussion in \cite{Earman:Lambda}, or also   \cite{Demianski:Cosmoconstant}.}

 In the present   use of  F-L models,  the cosmological constant is interpreted as (constant) vacuum energy density. Their  model space is usually parametrized by $(\Omega _m, \Omega _{\Lambda })$, the relative contributions of mass and of the cosmological constant (``vacuum'') to energy density.
The curvature contribution $\Omega _{\kappa }$ is related to the other two by $\Omega _m + \Omega _{\Lambda } + \Omega _{\kappa } = 1 $. In the Weyl geometric approach,  $ \Omega _{\Lambda }$ is a structural effect of the Weylian cosmological term, as we will see in remark \ref{Omega_m}, and contains no free  parameter. Weyl universes do  without the symbolical artefact of  an adaptable ``vacuum energy density'' which causes practitioners of cosmology so much headache, because it looks as if it were time-dependent (``dynamical''). 
 Some remarks on such questions will be made  in the outlook of this paper; a little more can be found in \cite{Scholz:arXiv}.
}
\end{remark}

The affine (W-L-C) connection of a Weyl universe can be calculated by Weyl's theorem 
(\ref{W-L-C}).\footnote{Here, as in other passages where it is adequate, we use the notational conventions of upper and lower indexes for tangent vectors, their linear coefficients, and their duals.} If we denote the coefficients of the affine (Levi-Civita) connection on the spatial fibre $S_{\kappa}$ by $\tilde{\Gamma}$ and use indexes $\alpha ,\beta ,\gamma = 1, 2, 3 $ , we find
\[ \Gamma _{\alpha 0}^{\alpha } = H , \; \Gamma _{\alpha \alpha }^{0 } = - g_{\alpha \alpha }H ,\;\; 
\Gamma _{\alpha \beta }^{\gamma  } = \tilde{\Gamma }_{\alpha \beta }^{\gamma }.
\]
Denoting the  covariant derivative  with respect to the Levi-Civita connection in $S_{\kappa}$  in direction $Y$ by  $\tilde{D}_Y$ we get for the  
covariant derivatives in $M$  for vector fields  $Y,Z$  tangential to the spatial fibres
\[  D_{\partial _0}  \partial _0 = 0, \; \; \;  D_{\partial _0}  Y = H Y , \;\; \;
D_Y Z = - g(Y,Z)H + \tilde{D}_Y Z .\]
With these expressions we calculate the curvature quantities we are interested in:

\begin{proposition} \label{curvature lemma}
A Weyl universe,   $M = \R \times S$, with Weyl gauge $(g, \varphi )$, spatial fibre $S := S_{\kappa}$  of ex-ante sectional curvature of space sections $\kappa$, and  $\varphi = H d x^0$ has  Ricci curvature 
\[  Ric = 2 (\kappa + H^2) d\sigma ^2 = - 2 (\kappa + H^2) g_{\alpha \alpha } (dx^{\alpha }) ^2 .\]
It has an effective sectional curvature 
 in 2-directions tangential to  $S$ (i.e., sectional curvature of in direction of $S$ embedded in to $M$)
\[  \kappa _S = \kappa + H^2\]
and scalar curvature
\[  \bar{R} = - 6 ( \kappa + H^2) . \]
Components, not stated  explicitly, of  curvature quantities  here given,   are 0 (e.g., sectional curvatures containing a timelike direction, etc.).
\end{proposition}

If doubts should have been left, the expression for the scalar curvature shows again, that the defining gauge of $M$ is  Weyl gauge.

As a {\bf corollary}, we calculate  the Einstein tensor, \footnote{The note \cite{Scholz:arXiv} contains  a sign error in the $d \sigma ^2 $ - term}

\beq \label{l-h-s} Ric  - \frac{1}{2} \bar{R}g =  3 ( \kappa + H^2) dt^2   -   ( \kappa + H^2) d\sigma^2  , \eeq
and  the Weylian cosmological term defined in equation (\ref{Weyl-Lambda}) 
\beq \label{cosmo-term}  \Lambda = 3  H^2 dt^2   -   H^2 d\sigma^2 .   \eeq
Thus the source term of the Einstein equation has the form of an energy-momentum tensor of an ideal fluid,
\beq  \label{fluid}   8 \pi \left[ \frac{G}{c^4} \right]   T = 8 \pi   \left[ \frac{G}{c^4} \right]   (\rho dt^ 2 + p d\sigma ^2 ) \eeq
(square bracket terms for values in physical units, in geometrical units $c = G = 1$), although  with  negative (repulsive) pressure. The general form is no surprise,  as in any Robertson Walker manifold the r.h.s. of the Einstein equation acquires the form of the energy-momentum of  a ``strange'' fluid  (\cite{ONeill}). The important, and from the standard view surprising point is that 
 constant values for energy density and pressure can go in hand with redshift. That justifies the announcement of this result as:
 \begin{theorem}
The source term of the Einstein equation of a Weyl universe with ex-ante sectional curvature $\kappa $  and Hubble constant $H$ is an energy-momentum tensor of a  strange \footnote{``Strange''  fluid, because of negative pressure.}
 ideal fluid    (equ. (\ref{fluid})).  In  gravitational  gauge (Weyl gauge), energy density $\rho $ and pressure $p$ are constant  and have the values
\[  \rho = \frac{3}{8 \pi}  \left[  \frac{c^4}{G} \right]  ( \kappa + H^2) , \;\;\; p =    -   \frac{1}{8 \pi } \left[  \frac{c^4}{G} \right]  ( \kappa + H^2) \, . \]
\end{theorem}

\begin{remark} \label{Omega_m} {\em 
The crucial difference between  the  statical models of cosmology and Weyl universes  is expressed by the Hubble form and the Weylian cosmological term (\ref{cosmo-term}). If these are  more than  formal gadgets, they represent the geometrical aspects of a tired light/gravitational modification of the transfer of wave-vectors along null geodesics (compare remark 
\ref{remark wave-vector}). In such a case, one can reasonably expect that the matter-field Lagrangian ${\cal L}_M$ which gives rise to the  the r.h.s. of the Einstein equation (\ref{r.h.s.}) splits into a matter term ${\cal L}_m$ proper and a term expressing the tired light ``mechanism'' (which may very well be quantum physical) ${\cal L}_{\Lambda}$,
\[  {\cal L}_M  = {\cal L}_m + {\cal L}_{\Lambda } . \] In  analogy to how standard cosmology interprets its cosmological term as ``vacuum energy'', we can then split up the r.h.s. of the Einstein equation of Weyl universes  into a ``mass term'' $\rho _m$ and a  ``cosmological term'' $\rho _{\Lambda}$ of  the energy density: 
\[   8 \pi \left[ \frac{G}{c^4} \right]   \rho _ m = 3 \kappa  \;\;\; \mbox{and} \;\; \;
 8 \pi \left[ \frac{G}{c^4} \right]   \rho _{\Lambda} = 3 H^2 \, .  \] 
It is important to keep in mind that the Weylian cosmological term {\em cannot be considered as ``vacuum contribution''} in the sense of the standard approach, as it is not of the form $const \cdot g$. Its contribution to the r.h.s. of the Einstein equation  has  the form of a strange fluid,  like the matter term  proper. It could therefore, in principle, just as well be assimilated to the latter. That also allows to consider the possibility that  only the energy density of the Weylian cosmological term is due to a physical structure lying at the base of the Hubble redshift (if different from the matter content of the universe), while the pressure term makes sense only in the overall balance ($p = p_m + p_{\Lambda }$).

If we use the standard definition of critical energy density,
 which  also for  Weyl universes  corresponds  to $ \kappa =0$,
\[ \rho _{crit} = \frac{3}{8 \pi}  \left[ \frac{c^4}{G} \right]  H^2  ,\]
 the relative contributions of the mass term $\Omega _m := \frac{\rho _m}{\rho _{crit} }$ and of the cosmological term $\Omega _{\Lambda }:= \frac{\rho _{\Lambda }}{\rho _{crit} }$  become 
\[   \Omega _m = \frac{\kappa }{H^2} = \zeta \; \;\;\; \mbox{and} \;\;\;
 \Omega _{\Lambda }  = \frac{H^2 }{H^2} = 1 \, . \]
These relations are just another expression of the beautiful correspondences between  sectional curvature values in spatial directions and energy densities, which have already been established. We rephrase them here  in geometrical units:
\[ 8\pi \rho = 3  \kappa _S  , \;\;\; 8\pi \rho _m = 3 \kappa , \;\;\;   8 \pi   \rho _{\Lambda} = 3 H^2 \, .\]
That is, the total energy density determines the ex-post sectional curvature in spatial directions of a Weyl universe, matter density determines the ex-ante curvature, and the energy density of the Weylian cosmological term determines the curvature contribution of the Hubble connection.
}
\end{remark}

\begin{remark}\label{remark hyle} {\em
The source term of Weyl universes satisfies the state  equation 
\beq  \label{hyle} p = - \frac{1}{3} \rho .   \eeq
At  first glance one may have  the impression that the   equilibrium problem raised by Eddington for the Einstein universe,  or even worse Einstein's gravitational collapse problem, might reappear in a slightly modified form. This is not the case, however, as Weyl universes do not rely on the classical cosmological term. Moreover, 
 we now know   that star and galaxy  matter  (luminous matter $\rho _{lum}$ with proportionality factor $\Omega_{lum} = \frac{\rho _{lum}}{\rho _{crit}} \approx  0.005 $) makes up only a very small part of the matter content of the universe. The main bulk of material content of the universe seems to be ionized gas, plasma  (most of it ionized hydrogen H$_1$), mixed with some neutral gas and dust components. Internal  repulsion  in the intergalactic medium, arising from electromagnetic forces  and, perhaps,  elementary particle physics  components, have to be taken into account. Their overall action may very well   balance  Einstein's gravitational attraction which, taken for itself, would let  ``the world'' appear  to be   damned to collapse. 

In our context, there is no reason for bringing exotic matter  into play.
 ``Exotic matter''  has  been  postulated as an
 ad-hoc hypothesis to avoid refutation of the theory of  primordial  nucleosynthesis by empirical data on dynamical mass density and to alleviate the problems arising in standard cosmological theories of  structure formation.  The dynamically determined mass density  $\rho _{d}$,  with relative value compared to the critical density $\Omega_{d} \approx 0.3$, is  now  at least one order  of magnitude higher  than the
 maximal density of baryonic matter, allowed by  the hypothesis of primordial mucleosynthesis,  $0.04 \leq \Omega_{bar} \leq 0.05$. No indications for exotic new types of matter have been found in several decades of intense research. It is  time to  accept   that the {\em primordial  nucleosynthesis hypothesis has been  refuted} in its core (baryon density) by empirical evidence, although certain features of it (high energy contribution to nucelosynthesis) will probably survive  in a modified way.\footnote{For a short hint, see outlook.} 

Thus we simply assume that dynamical  matter is dark (non-luminous) ordinary matter, dominated by  ionized gas, comparable to the one in  intergalactic space, only more concentrated around galaxies, clusters, and superclusters. Although the dynamics of the highly diluted ionized  matter  (apparently a  {\em plasma} in the sense of H. Alfv\'en, A. Peratt  e.a.)  has still to be understood, it is clear that the old game of contradicting attractive and repulsive  forces   comes back into the game, now as the interplay of gravitational and electrodynamical forces.\footnote{\label{footnote Peratt}For first interesting steps, see  \cite{Peratt:Plasmabook,Peratt:Plasma,Verschuur:Filaments}.} Cosmic space has to be  understood as   constituted in the interplay of such attractive and repulsive forces of matter. If Weyl universes turn out to be empirically satisfying models of cosmological observations on a broader range,  we even can {\em conclude} that there is dynamical equlibrium  in ``the cosmic mean'' (i.e., mean values formed over very large  regions of space-time), with equation (\ref{hyle})  as a global expression for such an equilibrium. The overall interplay of attraction and repulsion in the compound system of intergalactic plasma (including cosmic ray contributions), dust, luminous matter, etc.,  has to satisfy a state equation of a ``strange'' fluid. In allusion to an idea of Greek natural philosophers, I propose to call it the {\em hyle} equation. This term does {\em not impute}  new kinds of matter. 
}
\end{remark}

\subsubsection*{5. Redshift, light cone and luminosity decrease}
The defining gauge of Weyl universes is Hubble gauge, it is therefore immediate to write down the redshift function.   For a photon with trajectory $\gamma $, emitted  at $p = (t, x)$ and observed at $q = (t_0, x_0)$ ( $t< t_0$),  redshift is given, according to equ. (\ref{Hubble-z}), by
 \[ z (p,q) +1 =  e^{\int _t ^{t_0} \varphi (\gamma '(\tau )) d \tau } = e^{H_0 (t_0- t)} .\]
That agrees, of course, with the value expected from the Robertson-Walker picture (c.f. equs. (\ref{tau von x0}), (\ref{Rob Walk picture})),
\[  z+1 = \frac{H \tau _0}{H \tau } = e^{H_0 (t_0- t)}.  \]
Comparison with equ. (\ref{lambda-Weyl-universe}) shows
\[  z+1 = \lambda ^{-1} . \]
Here, the metrical perspective is completely different to the Robertson-Walker picture.   From  the perspective of Weyl  gauge, the   blow up of redshift, close to the  ``initial singularity'' of the Riemann gauge, appears  as the result of  a metrically deformed view of infinity. The same holds  for other quantities dependent on the cosmic time parameter.  As   Weyl gauge is  the  physically dominant perspective (matter gauge)   in our new models, with literally constant $G$ and constant energy densities, {\em this shift of perspective must not be dismissed as only a formal effect} of reparametrization,  
 
In particular, cosmic time distances, measured in  matter gauge  (more precisely, in the Riemannian component of the  Weyl metric of matter gauge) go to infinity with redshift:
\[   \Delta t := t_0 - t = H_0^{-1} \ln (z+1) . \]
Because it is such an important property, we formulate  the standardized form of the redshift development with time/distance as
\begin{proposition}
In a Weyl universe, the redshift-time relation for a photon emitted at $t < 0$  and observed  at  $t_0 = 0$  (``today'') is given by
\beq  \label{redshift-time}    z(t) = e^{H_0 |t|} -1 = \lambda ^{-1} \;\;\; \leftrightarrow \;\;\; |t| =  H_0^{-1} \ln (z+1) ,\eeq
where $\lambda $ is the scaling function from Weyl to Riemann gauge, normed by the condition $\lambda (0) = 1$.
\end{proposition}

From standard results on Robertson-Walker manifolds \cite[355f.]{ONeill} and by changing  coordinates as well as  gauge, one can determine explicit expressions for null geodesics in Weyl gauge. This allows to  ``verify'' (i.e. here, to check the consistency of definitions and methods) the relations between null geodesics, wave vector, length connection, and redshift given above in definition \ref{definition redshift} and  the equations of example \ref{remark wave-vector}. Here we content ourselves  with more  general observations on the light cone and its geometry.

Because of conformal invariance of the light cone, the {\em traces} of null geodesics  in Weyl universes with Weyl gauge $(g, \varphi)$ are identical to traces of null geodesics of the Riemannian component $g$ only. They can easily be described. Let us denote by $r$  the length of the projection of the segment $t \leq s \leq t_0$ of a  null geodesic $\gamma (s)$ into the spatial standard fibre $S_{\kappa }$ . Abbreviating the redshift between $t$ and $t_0$  by $z$, we find
\[  r= c |t - t_0| = H^{-1} \ln (z+1) . \]
We  call    the  intersection of the light cone with spatial fibres  a {\em light sphere}  and $r$ its  {\em radius}. The area of light spheres  and  volumes of  light cone segments can  easily be calculated.\footnote{Warning: In \cite[5]{Scholz:arXiv}  the area formula for Riemann gauge are given erroneously in place of of those in Weyl gauge. The magnitude formula on the same page is correct. }

\begin{lemma}\label{lemma light-spheres}
 In a Weyl universe of ex-ante sectional kurvature $\kappa $, Hubble constant $H$, and module $\zeta = \kappa H^{-2}$,   the following holds:  \\[0.8pt] \noindent
a) The area 
 $O = O_\kappa $ of   light spheres in dependence of redshift  is given by
\[  O_\kappa = \frac{4 \pi}{\kappa } SIN_k^2 (\sqrt{  \zeta } \ln (z+1)  \;  , \;\; \left( O_0= \frac{4 \pi}{H^2 } \ln ^2 (z+1)   \right) , \]
{\em where $k = \pm 1$ or $0$ and $SIN_1 := \sin$, $Sin_0 := id$, $SIN_{-1} := \sinh $.} \\[0.8pt] \noindent
b) The volume $V(z_1, z_2)$ scanned by the light cone between redshift values $z_1 \leq z_2$ is
\beq \label{volume}   V(z_1, z_2) = \int \limits_{z_1} ^{z_2} (z+1)^{-1} O(z) dz  .\eeq
\end{lemma}

The {\bf proof} of the first statement is obvious, as we calculate  areas of  spheres with  radius $r = H^{-1} \ln (z+1) $ in   the  3-geometries of constant curvature radius $a = \kappa ^{-\frac{1}{2}}$   and use (\ref{zeta}). For b),  we only need to integrate over  infinitesimal volume layers $O d x^0$ of the light cone, substitute  $d x^0 = c d |t|$, and use equation 
(\ref{redshift-time}). \\[1.5pt]

The decrease of luminosity of (electromagnetic) radiation  sources with redshift results from a damping of the energy flow by two reasons, the change with increasing $z$ of the area of light spheres (``normally'', but not always,  an increase), and a damping of energy quantities due to length transfer of energy flux, or more precisely of calibration by transfer, in Weyl geometry.  In the standard expanding models, the latter effect is  interpreted as a reduction of flux etc. because of expansion of spatial sections.
Here we use a natural principle for the application of scale gauge to energy quantities:

\begin{principle}
Propagation of energy quantities of gauge weight $k$ in the ``cosmic ether'', i.e.,  the compound system of the gravitational and the electromagnetic fields, is mathematically represented by calibration transfer of gauge weight $k$ with respect to Hubble gauge.
\end{principle}

\begin{example}
{\em a) Photon energy $E$, respectively the wave vector $k$ is a scalar Weyl field of gauge weight $[[E]] = -1$ and propagates along null geodesics  by calibration transfer of weight $-1$ {\em in Hubble gauge} (see remark \ref{remark wave-vector}). This is no surprise; we have defined Hubble gauge with this property in mind. \\[0.5pt]
b)  The energy flux $F$ of radiation of astronomical objects is observed as energy per time and area. Thus $F$ has the gauge weight $[[F]] = [[E]][[t]]^{-1}[[l]]^{-2}$ $ = - 1- 1-2  = - 4$. According to the principle, it  propagates  by calibration transfer of weight $-4$.

 This is a  natural assumption in Weyl geometry. One may  compare it to the view of 
 standard cosmology, where  a damping factor $(1+z)^{-2}$ is assumed  for the energy flux, one factor $(1+z)^{-1}$ for redshift, another one due to time dilation for comoving observers. Surface brightness and intensity are  calculated with a damping factor $(1+z)^{-4}$ 
\cite[152f.]{Peterson:Quasars}. 
}
\end{example}

On this basis we can calculate the magnitude-redshift dependence in Weyl universes. 

\begin{theorem}
Assuming principle 1 for the propagation and measurement of electromagnetic energy in a Weyl universe, 
\begin{itemize}
\item[a)]
 the energy flux  $F(z)$  of a radiating cosmic source depends on  redshift $z$ and the area of light spheres $O(z)$ measured  in Weyl gauge like 
\[  F(z)  \sim (z+1)^{-4}   O(z) ^{-1} .  \]
\item[b)] After redshift  correction of the measured energy by the factor $(z+1)$, the corrected energy flux $F_{corr}$  is proportional to
\beq  F_{corr}  \sim (z+1)^{-3}   O(z) ^{-1}   .\eeq
\end{itemize} 
 \end{theorem}

\noindent {\bf Proof} \quad
a) We consider transmission of radiation from the source at a point $p = (t,x)$ to a point at $p_0 =(t_0, x_0)$ along a null geodesic $\gamma $.
According to principle 1 energy flux $F$ propagates by calibration transfer of gauge weight $-4$.  This damping of the energy  is   only  due to the effect of the transfer according to the Hubble form, respectively the gauge factor $\lambda $ from Weyl to Riemann gauge as its its integrated version. Here we consider  calibration depending on the origin of the path rather than on its endpoint; thus we have to take the reciprocal (see equ. \ref{calibration-transfer}), and  the flux  is  proportional to $\lambda ^4$,
 \[F(\gamma (t)) \sim    \lambda ^{4}(\gamma (t)) . \]
From equ. (\ref{redshift-time}) we know $\lambda = (z+1)^{-1}$; therefore the contribution of the calibration transfer to damping is
\[ F(z)  \sim   (z+1)^{-4} .  \] 
In addition,  radiation energy emitted at $(t,x)$ during a short time interval  (an ``infinitely short'' one)  is distributed over the whole light sphere at the spatial section  lying over $t_0$, with area $O(z(t,t_0)) =: O(z)$. That results in an area contribution to damping according to 
\[  F(z)  \sim  O(z)^{-1} .\]
If we neglect  absorption by cosmic matter, we arrive at an overall  effect in Weyl geometry 
\[ F(z)   \sim   (z+1)^{-4}   O(z) ^{-1} . \]
The correction in b) is direct. \\[1pt]

\begin{remark} {\em 
Readers who are doubtful of principle 1,  may want  to avoid  calibration by transfer and prefer to  calculate  in the Robertson Walker picture 
 with traditional methods of scaling down the flux in  ``expanding universes'' (not to forget the gauge factors between Weyl and Riemann gauge).  The result should be the same.  In any case, the empirical data of supernovae Ia magnitudes are a beautiful empirical test for the physical  acceptability of the principle {\em in the frame} our approach (see below, outlook).
}
\end{remark}

 In the sequel we    refer  to the redshift corrected values, $F_{corr}$, if not explicitly stated otherwise, because the redshift-energy correction as in part b) of the theorem is usually applied by astronomers  during the $K$-correction for   apparent magnitudes \cite[80]{Hubert:Kosmologie}.

The apparent magnitude of a cosmic source is defined by
\beq \label{mag-1} m := - 2.5 \log F  + C , \eeq
where the constant $C$ contains the dependence on the {\em absolute magnitude} $M$, i.e., the magnitude in which  the same source would appear to a terrestrial observer, if it was placed at the     norm distance of classical astronomy $d_0 := 10 $pc = $10^{-5}$Mpc (just in reach of good parallax measurements). Moreover,  $C$ depends on the cosmological assumptions in each model, in particular on the value for the Hubble constant $H_0$. 
With  the values for $F$  from the last theorem we get
\[  \label{mag-Weyl}  m = - 2.5 \log (z+1)^{-3} O(z)^ {-1}  + C = 5 \log (z+1)^{\frac{3}{2}} \sqrt{O(z)} + C  ,  \]
and with  lemma \ref{lemma light-spheres}:\\[2pt] \noindent
{\bf Corollary} \quad {\em
In a Weyl universe with module $\zeta $  and spatial ex-ante curvature  type $k= -1, 0, +1$,  the 
 redshift-corrected apparent luminosity $m_k $ of a cosmic radiation source is: }
\beq \label{luminosity1}
m_k(z)  = 5 \log \left( \frac{  (z+1)^{\frac{3}{2}}}{ \sqrt{| \zeta |}} SIN_k( \sqrt{| \zeta |}   \ln (z+1) ) \right) + C ,   
\eeq
($SIN_1 = \sin$, $Sin_0 = id$,  $SIN_{-1} = \sinh$, as above). {\em  For $k=0$, this is to be understood in the sense of  the limit $\zeta \rightarrow 0$, }
\[ m_0 = 5 \log  \left( (z+1)^{\frac{3}{2}}   \ln (z+1) ) \right) + C .     \]
\\[1.5pt]
 
Sometimes it is  useful to   substitute  the constant $C$ in the formulae of the corollary by  an expression in the absolute magnitude at redshift   $z_0 = e^{H d_0} - 1 \approx  H\, d_0$,
\[ C = M -   5 \log \left( \frac{e^{1.5 \, H d_0}}{ \sqrt{\zeta }} SIN( \sqrt{\zeta }  H d_0)\right) \approx  M - 5 \log (H d_0) .  \]
 Then we get an explicit  function for
$  m_k(z, M, \zeta , H)$, depending  on the redshift, the absolute magnitude,  the module $\zeta $ of the Weyl universe (or $\Omega _m$), and the Hubble constant. The last approximation is so good that it can be taken as a ``numerical equality'' for all practical purposes.

 If we accept the source term interpretation of the Einstein equation as in remark \ref{Omega_m}, $\zeta = \Omega _m$, we get a ``more physical'' version of (\ref{luminosity1}). For $k=1$,  it acquires the form
\beq  \label{luminosity2}
  m_k(z)  = 5 \log \left( \frac{  (z+1)^{\frac{3}{2}}}{ \sqrt{ \Omega _m}} \sin ( \sqrt{ \Omega _m }   \ln (z+1) ) \right) + C \, . \eeq

The great simplicity of Weyl universes results in simple  explicit expressions for  their 
geometrico-physical properties. As a last example,  the angular size of objects with diameter $b$ in  distance $d$ can easily be calculated  from spherical trigonometry  for $k=1$,  because   the light cone structure is  conformally invariant  under deformation of the classical Einstein universe to the Einstein -Weyl one. Angular sizes of such objects  are   given by 
\beq \label{angular-size} \sin \frac{ \alpha  }{2 } = \frac{ \sin\frac{ b }{2 a } }{\sin \frac{ d }{ a} } ,\eeq
where $a$ is the ex-ante radius of curvature of spatial sections, $a = \kappa ^{-\frac{1}{2}}= \frac{H^{-1}}{ \sqrt{\zeta } }$.

\subsubsection*{6. Outlook } 
Before we come to the end,  we want at least  to indicate how well the new models  behave with respect to observational data.
We start with the recent {\em  luminosity-redshift measurements of supernovae Ia } by the {\em Cerro Tololo Inter-American Observatory} group (CTO)   and  the {\em Supernova Cosmology Project} (SCP)   \cite{Perlmutter:SNIa}. These data, complemented by those of the group around A. Riess,   have been interpreted as   striking  evidence in favour of the Friedman-Lema\^itre model with non-vanishing  cosmological constant \cite{Groen:Universe}. Together with constraints derived from the ``inflationary'' hypothesis they lead to the presently accepted standard density parameters  $(\Omega _m, \Omega _{\Lambda }) \approx  (0.3, 0.7)$.
It is therefore quite  remarkable that in 
 our  framework {\em  the supernovae data can be fitted just as well by Weyl universes } (figure 1). 
With   $H \approx  2.3 \;10^{-4} \, Mpc^{-1}$ ($H_0 \approx  70 \, km s^{-1} Mpc^{-1}$) and   $M \approx -19.3$ for the absolute magnitude of supernovae Ia,
 the best fit for the module of Weyl universes is 
\[  \zeta \approx 1  \, , \; \;\;  \mbox{ with residual dispersion} \;\;  \sigma \approx  0.318 \approx 1.24 \sigma _{data}.\footnote{The fit   quality for the  standard model is $\sigma_{stand} \approx  0.309$ for the parameter choice $(\Omega _m, \Omega _{\Lambda }) = (0.3, 0.7)\approx  1.21 \sigma _{data} $  (my calculation, E.S.). The difference is not significant.}
 \]
 This  result can be considered as a first  empirical answer to the question, whether the hypothesis of luminosity decrease according to calibration by transfer in Weyl geometry (principle 1) makes sense or not. {\em It clearly does.}

\unitlength1cm
\begin{figure}[bottom]
\center{\includegraphics*[height=5.0cm, width=9.0cm]{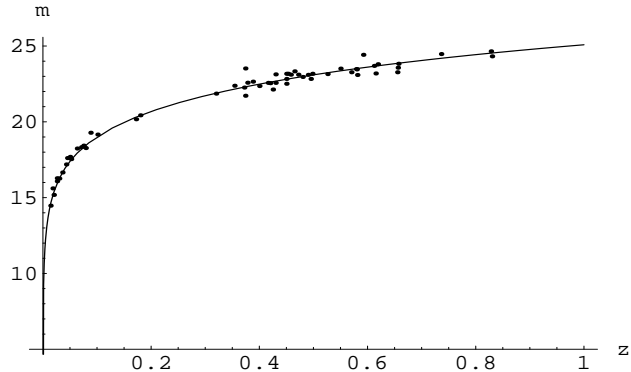}}
\caption{Magnitude-redshift relation in  Einstein-Weyl universe, $\zeta = 1$ (no graphical distinction to  $\zeta = 1.5$), $H_0 = 70$, $M= - 19.3$, compared with recent SN$_{Ia}$ data of the SCP and CTO groups (dots)  }
\end{figure}

The fit quality remains very stable under change of  $\zeta$. If we relax only  a little bit to  $\sigma \leq 1.3 \,  \sigma_{data}$, we  get a wide interval for admissible curvature parameters, including all three curvature types $k= \pm 1, 0$
\[  \zeta \in [ -0.1, 3.6 ]=: I_{SNIa} .\]
 For  $\zeta =0.3$,  the residual dispersion is still only  $\sigma _{0.3} \approx 0.325$.  Because of  the possible identification    $ \Omega _m = \zeta $  (remark \ref{Omega_m}), this  value   is of particular interest, if we want to adapt the model to present data of dynamical mass density. From the empirical point of view, the  supernovae luminosities and the mass density criterion clearly single out  the {\em Einstein-Weyl universes}, $k= +1$, among the whole model class.

Other important data  for testing cosmological models are  {\em quasar frequencies } in dependence of redshift.
The best values presently  available  are those of the {\em Sloan Digital Sky Survey} (first data release)  \cite{SDSS:2003b}.
The colleagues of the SDSS have measured  16 713 objects up to $z \leq 5.4$ and gave a concentrated survey  in a histogram with  69 intervals of width $\Delta z = 0.076$. The frequencies show a clear and 
well-formed increase up to about $z\approx 2$ and a rapid decrease for $z > 2.1$.
It is very interesting to see  that the peak of observed quasar frequencies about $z \approx 2$ corresponds closely to the maximum in volume scanned by the light cone  (in equal  $z$-intervals) in Einstein-Weyl universes (equ. (\ref{volume})).\footnote{Because of the inflationary dogma ($k=0$) and the expanding geometry, this effect is suppressed   in the standard approach.}

A calculation of the numbers of objects to be expected under the assumption of equal distribution in Einstein-Weyl universes for $\zeta = 0.3, \, 0.8 , \, 1, \, 1.5$ and $2$ gives   a surprisingly good agreement of the geometry of the light cone with the observed quasar frequencies up to the peak at $z \approx 2$. For larger values of $z$,  observed quasar frequencies  break down  too  abruptly to be accounted for by  geometrical reasons only (figure 2).\footnote{Densities of objects have been calculated  from respective volumes in Einstein-Weyl universes and SDSS data for $z \leq2.24$, shortly after the peak.}
In fact the rapid decline of observed quasar numbers  is  strongly enforced by  the reduction of sensibility of the CCD detectors for $z > 2$ and, perhaps,   other selection effects. 

\begin{figure}[h]
\center{\includegraphics*[height=7.5cm, width= 10cm]{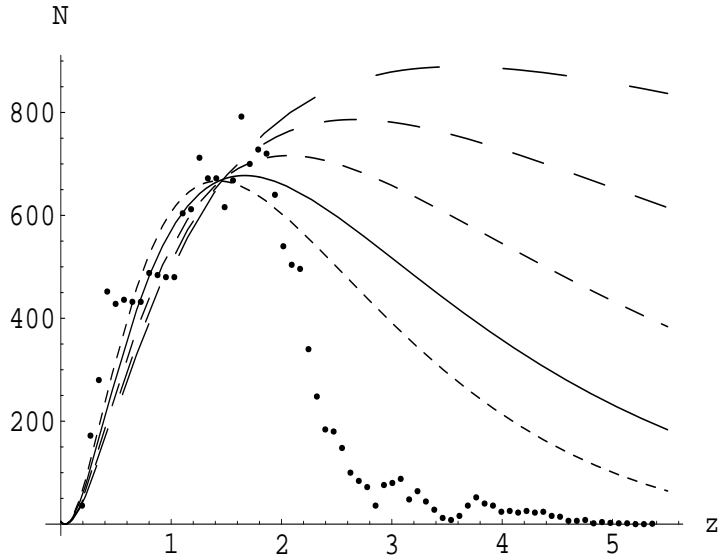}}
\caption{Frequencies of equally distributed objects in Einstein-Weyl universe with $\zeta =0.3$ (coarsest dashing), $\zeta =0.6$ (coarse dashing),  $\zeta =1 $ (medium dashing), $\zeta =1.5$ (undashed),  and $\zeta =2$ (finest dashing), compared  with quasar frequencies of SDSS (dots); interval  width $\Delta z = 0.076$  }
\end{figure}

Already a rough comparison of the data shows  that a   parameter  interval 
\[ \zeta \in I_{QF} :=   [0.3, 2] \]
 fits in with  quasar frequencies up to $z \approx 2$. For larger values of $z$, it leads to an acceptable agreement only for $\zeta \geq 0.6$ and if sufficiently strong selection effects are at work. Which range of the interval $I_{QF}$ is to be preferred,  depends  on an estimation of selection effects and a comparison of the reliability of other data which are taken into account.

The determination of {\em  dynamical mass density} close to galaxies, clusters, and superclusters leads to an actual estimation of 
$ 0.1 \leq \Omega_m \leq 0.5  $ for smaller structures and of $ 0.2 \leq \Omega_m  \leq 1 $ for larger ones,  with a preferred value close to  $ \Omega_m \approx  0.3 $ \cite{Carroll:Constant}. 
That is consistent  with  supernovae and quasar data, but probably not the best we can get. We may reasonably expect a rising value for  $ \Omega_m$ with increasing   awareness for dynamical mass in larger regions. 

Mass density estimations are precarious data. The best measured cosmological data are  those of the {\em cosmic microwave background} (CMB) with its  precise Planck profile and a radiation temperature $T_{CMB} \approx 2.726\; ° K $ with error less than $0.1 \%$ (after subtraction of the dipole moment of motion against the CMB rest system). 

 In
Einstein-Weyl universes,  a  beautiful alternative to the standard  explanation of the CMB (as the redshifted view upon a ``surface of last scattering'') is possible by a theorem of I.E. Segal.  Segal has proved that  the quantized Maxwell field on the Einstein universe has a (maximal entropy) equilibrium state with radiation  of exact Planck characteristic  \cite{Segal:CMB1}. The proof     translates to the Weyl case because of conformal invariance of the field equations. It is now the question, whether (or how) the Segal background is related to the cosmic background radiation.

  In an Einstein-Weyl universe, the Segal background plays the role of a vacuum ground state for the electromagnetic field.  We may thus reasonably {\em conjecture}: 

($\ast$)  {\em The energy density $ \rho _{\Lambda} $ of the Weylian cosmological term is due to the  energy density $\epsilon _{Segal}$ of the Segal background, $   \rho _{\Lambda} =  \epsilon _{Segal} $.} \\[-3.5mm]

If this is true, $\epsilon _{Segal}$ is equal to the  critical  density  (remark \ref{Omega_m}),  
$  \rho _{crit} \approx 5.2 \, keV \, cm^{-3}$,
  and corresponds to a Planck   temperature $T_{Segal} \approx 32\, °K$. Although that may  appear   unlikely at first glance,  a second thought shows  that any direct identification of such a background equilibrium state will be difficult or even impossible. Everything, including each part of the measuring device, will be immersed in a diffuse bath of the $32\, °K$ radiation, which suppresses its discernibility as incoming radiation. Moroever, its spectral range  lies deep inside the  interval of the cosmic infrared background which is very difficult to measure anyhow \cite{Dwek/Hauser:2001}. 
Thus there seems to be no chance for a {\em direct} empirical identification of
the  emission component due to   the 
Segal background.

\begin{figure}[h]
\center{\includegraphics*[height=5.5cm, width= 10cm]{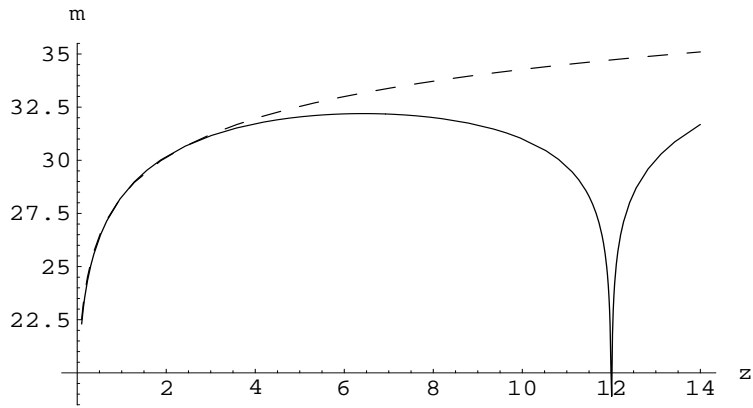}}
\caption{Refocussing of luminosity (here for a source of absolute magnitude $M=-16$) in Einstein-Weyl universe with $\zeta =1.5$ (undashed) at the first conjugate point, $z\approx 12$,  in comparison with luminosity of an equally strong source in the standard cosmological model ($\Omega _m, \Omega _\Lambda ) = (0.3,0.7)$) (dashed) }
\end{figure}

 On the other hand,  the radiation characteristics of the region close to the conjugate point of a general observer in  an   Einstein-Weyl universe would be strongly dominated by the (redshifted) Segal background. Therefore  the signal from  this region would have a clearly discernible  Planck characteristic,  be nearly ideally isotropic, and would be   highly refocussed at the point of observation (see figure 3). A sharp delimitation of refocussing  about  the redshift at the conjugate point suppresses a smearing out of the Planck characteristic of the source, which would otherwise happen. The signal will  be  corrupted only by interaction along the passage through the spatial sphere $S_{\kappa }$ with hot gas around clusters and superclusters, or a slightly increased energy density of the Segal background in these regions.
We   have to look at empirical indicators which allow to decide whether the conjecture ($\ast$) and its consequences are realistic or not. 

If the CMB is due to emission  of the Segal background  in the vicinity of ``our'' conjugate point,
an easy calculation shows that
  the redshift at the conjugate point has to  be
\[ z_{conj} = \frac{T_{Segal}}{T_{CMB}}  \approx \frac{32}{2.73}\approx 12.\]
 That corresponds to a parameter value $\zeta \approx 1.5 \, \pm 5 \%$ of the Einstein-Weyl universe and {\em stands    in pleasing agreeement with quasar frequencies and SNIa luminosities.} We may  consider this agreement as a first positive empirical test for the conjecture. Of course more  of them are needed, most important a detailed   analysis of anisotropy data from this point of view, before we can claim more than a hypothetical plausibility for it.

On the theoretical level, it is worthwhile  to notice that, in the frame of Weyl universes and under the assumption of the correctness of ($\ast$),   {\em the two empirically most important cosmological parameters,} temperature $T_{CMB}$ of the microwave background and Hubble constant $H$, {\em determine $\zeta $ and with it the physical mean geometry of the cosmos in the large without ambiguity, in principle, and with high precision in practice.}\footnote{ The small error bound $\pm 5 \%$ for the estimation of $\zeta $ arises from the fourth root to be drawn from the energy density and thus from $H^2$.}
 If other observations support the conjecture, we will attribute  these data  much more credibility than  mass density estimations from dynamical observations, which cannot account for  the diluted mass in  intercluster regions anyhow.

If the Segal background  gives  a realistic explanation of the cosmic microwave background (with or without conjecture ($\ast$)), the 
anisotropies arise  naturally as a  result of the passage through material inhomogeneities, in particular  cosmic plasma (``hot gas'')  concentrated around  clusters, cluster groups, and superclusters, or through regions of slightly increased temperature of the Segal background.  The most  important contributions to  anisotropies  with multipole moment $l \approx 200$  are  expected to be caused by objects on the  supercluster level in the segment of the spatial $S_{\kappa }$ with $1 \leq z \leq 6$, which covers $80 \, \%$ of the total volume until the first conjugate point. For $\zeta = 1.5$, objects of diameter $40 \; Mpc$ in this redshift range appear under an angular size between $0.7\, \deg$ and $1 \, \deg$ (figure 4).
\begin{figure}[h]
\center{\includegraphics*[height=5.5cm, width= 10cm]{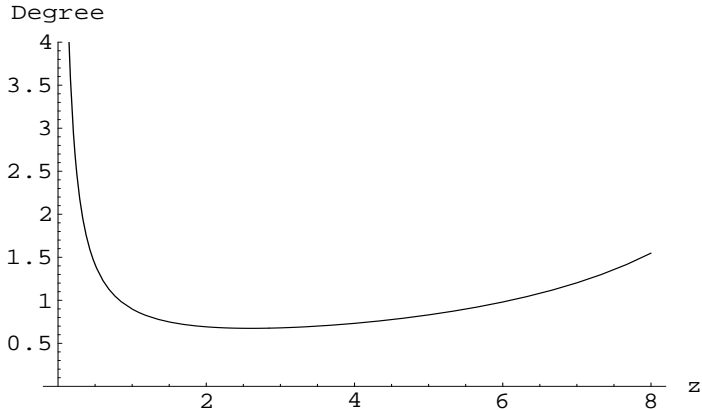}}
\caption{Angular size of objects with diameter $40\,  Mpc$ in Weyl universe with $\zeta = 1.5$ }
\end{figure}
and lead to a peak of the anisotropy signal at a multipole moment of about $l \approx \frac{180}{0.8} = 225$.

It is very  encouraging that astronomers have started to investigate the possibility of an influence of clusters and higher order structures on CMB anisotropy. Recent evaluations of  WMAP anisotropy data  by a group around T. Shanks \cite{Shanks:WMAP} have demonstrated that there exists, in fact,   a {\em strong 
anti-correlation between temperature and cluster positions for redshifts up to  $z\approx 0.2$.} The authors explain their observation by inverse Compton scattering of photons  in hot gas in  clusters and cluster groups (``Sunyaev-Zeldovich effect''). They  consider their results as a  corruption of a more original anisotropy signal stipulated in the standard picture to result from some legendary "accoustic waves'' in the primordial soup at the end  of the ``inflationary phase''. Shanks e.a.  expect that with a continuation of the evaluation to $z\approx 0.5$  the ``contamination'' of anisotropy data, as they call it, may rise considerably.

From our point of view, that {\em   has} to happen. We even  can quantify the degree of ``contamination'', which is to be expected. It is closely related to the volume increase covered by  the light cone in respective $z$-intervals. The interval $0 \leq z \leq 0.2$ covers $0.23\,\%$ of the total volume (up to the first conjugate point) in a Weyl universe of $\zeta = 1.5$. Contributions to the anisotropy signal with $l \approx 200$ from redshifts $0.1 \leq z \leq 0.2$ are caused by objects on the cluster level (up to $10 \, Mpc$). The main anisotropy signal is expected to be caused by superclusters with $1 \leq z \leq 6$, as indicated above. Taking the number of partial systems (clusters  in superclusters) into account, the anisotropy contribution of this redshift range should be an order of magnitude larger than its volume ratio. Thus it should account for about 1 -- 3 $ \, \%$ of the whole anisotropy signal of the CMB.
For the next interval, $0.2 \leq z \leq 0.5$, envisaged by the group around T. Shanks, the volume proportion rises to $2.5 \, \%$. Anisotropies with $l \approx 200$ result here from objects in cluster group size. We therefore expect a contribution to CMB anisotropies of about $10 \, \%$. 

It would be very welcome if the  estimation of the  anisotropy contributions could be continued until $z \leq 1.5$ or even $z \leq 2$.  For $z> 0.8$, structures on the supercluster level come into the play.  For these redshifts, our  volume estimations let us expect  a contribution to the whole anisotropy signal of $35 \, \%$ and $43 \, \%$,  respectively. 
 A detailed investigation of the CMB anisotropies may even be able to shed  light on the origin of the   ``tired light mechanism'', if this is in fact the reason for  the good modelling behaviour of Weyl universes.
The reduction of photon energy during the passage of clusters (or higher order structures) may have the same general reason as the Hubble  redshift itself.  A photon-photon interaction with the Segal background, slightly increased in energy/temperature in regions of luminous matter concentrations  may  be an additional cause (``Freundlich-Vigier-Segal effect''), or even an alternative,  to the Sunyaev-Zeldovich effect. In this case  we had traced a common   
 physical basis for both, the Hubble redshift and the tiny anisotropies impressed on the microwave background signal (or at least parts of them).

Other points have to be discussed with the necessary care and in greater detail than can be done here.  Most important among them are those  related to the observation that expansion  may only   be due to a   geometrical fiction of  the model class which has been accepted as {\em the} symbolical a-priori during the last decades  (F-L models).   If    physics of the cosmos is  better characterized by Weyl gauge,  expansion appears as a merely formal feature of  Riemann gauge. In fact,  in the standard approach the dependence of  $\Omega _{\Lambda }$ on the cosmic time parameter and the famous ``accelerated expansion'' of the universe suggest that all this may  just be a model artefact, without  direct  (realistic)  physical meaning.  The behaviour of the  ``dynamical dark energy''  in the standard approach  hints even more strongly  in this direction. 

The attempts of achieving a satisfying model of structure formation by an interplay of repulsive  dynamical dark energy plus cosmic expansion and gravitational attraction have  run into an impasse. Even under the auxiliary  assumption of fictitious gravitational material (exotic dark matter), it was only possible to imitate
 either  large scale structures (``hot dark matter'') or small scale structures (``cold dark matter''), but never both together \cite{Ostriker:Darkmatter}.  From these heroic attempts we can  draw the lesson that in the standard picture  the repulsive force is given a much too schematic expression to  allow for sufficient  flexibility, or even (structural) ``creativity'' of matter. We therefore  have reasons to look forward for progress in the plasma physical alternative program which tries to understand structure   formation  by an interplay of gravitational attraction with electromagnetic attraction and repulsion inside the ionized cosmic medium. That will surely be easier without the expansionary paradigm than within it.\footnote{Cf. footnote \ref{footnote Peratt}. It will be a  challenging  task for the history and philosophy  of  science to have a closer look at  the cultural conditions   which let it appear plausible to two generations of scientists  that 
 a paradigm of nuclear explosion, or of a  (post-) ``inflationary boost'', can serve as a  clue to  structure {\em formation} (rather than destruction) --- contrary to the modelling experience of several decades.  }

Metallicity values of galaxies and quasars give another occasion for a critical review of the idea of a global evolution correlated with expansion. Even though very detailed  and long range data are now available, they do not  show a systematic   correlation between redshift and metallicity \cite{Pagel:Nucleo}. Apparently, this lack is  not seen as a problem in the actual discussion.  To the surprise of astronomers and astrophysicists, however, the metallicity of the quasars with the highest   redshifts observed at  present  shows no indication  of lower overall values than quasars and galaxies closer by \cite{Corbin:QSOs}. Adding to the surprise,  very metal poor  galaxies of  small redshift have been detected \cite{Grebel:metal_poor}. {\em The metallicity of galaxies and quasars turns out  to depend much more on local conditions} (most importantly the mass of the object) {\em and regional evolution histories than on a ``global evolution''} clear signs of which are even missing. In the present discourse, the observational facts are  explained  by diverse exceptional clauses to a general rule which seems to exist  in the theoretical expectation of the standard approach only.\footnote{I thank E. Grebel for information on this point.}

Finally, the empirical indicators of physical states of the cosmos ``close to the initial singularity'' are no longer as convincing as they appeared  to be some decades ago. 
As mentioned above, the hypothesis  of {\em primordial nucleosynthesis} has  been factually   refuted  by the  observational data on dynamical mass and the lack of success for diverse attempts to find a substitute in strange (``exotic'') matter (remark \ref{remark hyle}). 
Recent high precision lattice gauge  calculations of the assumed electro-weak phase transition (which has been considered  as the clue to baryogenesis in the ``early universe''  in  the standard approach) complement this picture. The calculations demonstrate the existence of  a  {\em  theoretical upper bound} of $74 \;GeV$ for a Higgs-boson  mass compatible with 
electro-weak phase transition  \cite{Fodor:ew_phase_transition}. This bound is inconsistent with  the present  {\em  experimental  lower limit} of  $ 89.8\; GeV$ for the Higgs mass. Because of  reliable  error estimates for the theoretical and for the experimental result,  {\em  the combination of the standard model of elementary particle physics (SM) with the standard approach to cosmology  (SC) has  lead  to another anomaly for the hypothesis of a ``primordial phase of the world''.}\footnote{The authors of the study comment their result in clear words: ``This also means that the SM baryogenesis in the early Universe is ruled out'' \cite[24]{Fodor:ew_phase_transition}. By SM the authors refer to the standard model of elementary particle physics. They consider their result as an argument for the necessity for a physics beyond SM.}

With primordial nucleosynthesis and the  electroweak phase transition, two inportant features of a hypothetical early phase of the expanding world picture have shown to be wrong.  A distributed locus  for  high energy contributions to nucleosynthesis,   which has to exist  in  addition to stellar nucleosynthesis,  has  entered  observability in the form of active galactic nuclei and quasars. Empirical and theoretical knowledge on these structures is  growing rapidly.  We may expect that they  provide more realistic candidates for high energy  nucleosynthesis (beyond stellar processes) than the hypothetical primordial  phase  of the universe in the old  paradigm, although new features like a cracking of higher nuclei into light ones   enter here and  change the overall function of the process (``recycling'' of baryonic matter rather than its ``genesis''). The Weyl geometric approach provides an alternative without the necessity to invent  repair measures inside  the paradigm of a ``hot initial phase of the universe''. 

If, in in one way or another, the most important signatures of a global  evolution {\em of the cosmos as a whole} turn out to be unreliable, we have to  content ourselves  for scientific investigations with observing  local, or  better,  regional evolution of {\em substructures  in the cosmos} and  to understand them   the best we can.  The grand narrations of  {\em The Primordial Phase of  The World} can be handed over again   to the  myths of all the different cultures on our earth. This is no real loss, by the way; questions relating to the evolution of the universe as a whole have  always been a part of speculative natural philosophy and continue to be so. In  the extremes,  they even may  be  considered as  either ``meaningless''  or as an attempt to approach ``the mind of God'' etc., depending on taste and the metaphysical perspective we assume. Somewhere in between stand the most recent quantum cosmological speculations of late modernity.

Here, we  could only scratch at the surface of these  difficult and multifarious problems. 
For the moment, we have to leave it with   first glimpses  into a vast and challenging terrain. If Weyl universes appear useful and theoretically attractive  to other researchers, there will  be  better occasions for  more  detailed discussions of the open questions and, hopefully, some new and unexpected answers.  \\ [2pt]

\small
\noindent
{\bf Acknowledgements:} I have to thank many people from different fields for criticism and support in different stages of the research for this project.  A talk of P. Cartier in September 2001 was important to me to take up these investigations.  In the early stage,  H. Goenner and J.H. Eschenburg contributed critically to  conceptual questions; H. Arndt and P. Feuerstein gave  computational support.  During  the course of the development of the approach I profited from   remarks by  K.-H. Kampert, Z. Fodor, D. Lind, W. Rhode, A. Rendall, T.Scholz, E. Grebel, D. Puetzfeld and an anonymous referee of an earlier version submitted to   (and rejected by) {\em Annalen der Physik}. 
 Matthias Kreck was the first mathematician, who took the enterprise seriously; he invested   time and enthusiasm in  discussions on the mathematics and  supported me to find ways to improve and to communicate it to others. Among physicists, the discussions at Freiburg with D. Giulini and H. Roemer gave important hints to me and the interest of J. Renn, J. Stachel and G. Ellis provided a kind form of encouragement.  Outside the professional world, it was a particular pleasure for me  that  Sven Heinze, Sonja Heinze, and  Angelika Scheuffele  responded with sympathy and curiosity in  the questions dealt with here.

\bibliographystyle{apsr}
  \bibliography{a_litfile,a_litfile_2}

\end{document}